\documentclass[english,3p]{elsarticle}
\usepackage[T1]{fontenc}
\usepackage[latin9]{inputenc}
\usepackage{color}
\usepackage{babel}
\usepackage{array}
\usepackage{bm}
\usepackage{multirow}
\usepackage{amsmath}
\usepackage{graphicx}
\usepackage{pdfpages}
\usepackage[unicode=true,
 bookmarks=true,bookmarksnumbered=false,bookmarksopen=false,
 breaklinks=false,pdfborder={0 0 0},pdfborderstyle={},backref=false,colorlinks=true]
 {hyperref}

\makeatletter

\providecommand{\tabularnewline}{\\}

\biboptions{sort&compress}

\makeatletter
\providecommand{\doi}[1]{%
  \begingroup
    \let\bibinfo\@secondoftwo
    \urlstyle{rm}%
    \href{http://dx.doi.org/#1}{%
        \discretionary{}{}{}%
     \nolinkurl{#1}%
   }%
  \endgroup
}
\makeatother

\usepackage{xcolor}
\usepackage{textcomp}
\usepackage{hyphenat}
\usepackage{makecell} 
\usepackage[]{lmodern} 
\usepackage{listings}
\makeatother

\usepackage{listings}

\begin{document}

\title{\textsf{SpaceGroupIrep}: A package for irreducible representations
of space group}

\author[bit]{Gui-Bin Liu\corref{cor}}

\ead{gbliu@bit.edu.cn}

\cortext[cor]{Corresponding author}

\author[bit]{Miao Chu}

\author[buct,bit]{Zeying Zhang}

\author[bit]{Zhi-Ming Yu}

\author[bit]{Yugui Yao\corref{cor}}

\ead{ygyao@bit.edu.cn}

\address[bit]{Key Lab of advanced optoelectronic quantum architecture and measurement
(MOE), Beijing Key Lab of Nanophotonics \& Ultrafine Optoelectronic
Systems, and School of Physics, Beijing Institute of Technology, Beijing
100081, China}

\address[buct]{College of Mathematics and Physics, Beijing University of Chemical
Technology, Beijing 100029, China}
\begin{abstract}
We have developed a Mathematica program package \textsf{SpaceGroupIrep}
which is a database and tool set for irreducible representations (IRs)
of space group in BC convention, i.e. the convention used in the famous
book ``The mathematical theory of symmetry in solids'' by C. J.
Bradley \& A. P. Cracknell. Using this package, elements of any space
group, little group, Herring little group, or central extension of
little co-group can be easily obtained. This package can give not
only little-group (LG) IRs for any k-point but also space-group (SG)
IRs for any k-stars in intuitive table form, and both single-valued
and double-valued IRs are supported. This package can calculate the
decomposition of the direct product of SG IRs for any two k-stars.
This package can determine the LG IRs of Bloch states in energy bands
in BC convention and this works for any input primitive cell thanks
to its ability to convert any input cell to a cell in BC convention.
This package can also provide the correspondence of k-points and LG
IR labels between BCS (Bilbao Crystallographic Server) and BC conventions.
In a word, the package \textsf{SpaceGroupIrep} is very useful for
both study and research, e.g. for analyzing band topology or determining
selection rules.
\end{abstract}
\maketitle

\global\long\def\vr{\bm{r}}%
\global\long\def\vR{\bm{R}}%
\global\long\def\vk{\bm{k}}%
\global\long\def\vK{\bm{K}}%

\global\long\def\bktwo#1#2{\langle#1|#2\rangle}%

\global\long\def\bkthree#1#2#3{\langle#1|#2|#3\rangle}%

\global\long\def\ket#1{|#1\rangle}%
\global\long\def\bra#1{\langle#1|}%

\global\long\def\ave#1{\langle#1\rangle}%

\global\long\def\veps{\varepsilon}%

\global\long\def\herring#1{\vphantom{#1}^{H}\!#1}%

\section*{Program summary}

\noindent \textit{Program title}: \textsf{SpaceGroupIrep}

\noindent \textit{Developer's respository link}: \url{https://github.com/goodluck1982/SpaceGroupIrep}

\noindent\textit{Licensing provisions}: GNU General Public Licence
3.0

\noindent\textit{Distribution format}: tar.gz

\noindent\textit{Programming language}: Mathematica

\noindent\textit{Classification}: 11.2

\noindent\textit{External routines/libraries used}: \textsf{spglib}
(\url{http://spglib.github.io/spglib})

\noindent\textit{Nature of problem}: Space groups and their representations
are important mathematical language to describe symmetry in crystals.
The book\textemdash ``The mathematical theory of symmetry in solids''
by C. J. Bradley \& A. P. Cracknell (called the BC book)\textemdash is
highly influential because it contains not only systematic theory
but also detailed complete data of space groups and their representations.
The package \textsf{SpaceGroupIrep} digitizes these data in the BC
book and provides tens of functions to manipulate them, such as obtaining
group elements and calculating their multiplications, identifying
k-points, showing the character table of any little group, determining
the little-group (LG) irreducible representations (IRs) of energy
bands, and calculating the direct product of space-group (SG) IRs.
This package is a useful database and tool set for space groups and
their representations in BC convention.

\noindent\textit{Solution method}: The direct data in the BC book
is used to calculate the LG IRs for standard k-points defined in the
book. For a non-standard k-point, we first relate it to a standard
k-point by an element which makes the space group self-conjugate and
then calculate the LG IRs through the element. SG IRs are obtained
by calculating the induced representations of the corresponding LG
IRs. The full-group method based on double coset is used to calculate
the direct products of SG IRs. In addition, an external package \textsf{spglib}
is utilized to help convert any input cell to a cell in BC convention.

\section{Introduction}

Symmetry embodies the beauty of natural law and hence plays an important
role in physics. A famous example is the Noether's theorem which relates
symmetric invariance to conserved quantities. Another example is various
selection rules determined by symmetry. Furthermore, in recent years
point-group (PG) and space-group (SG) symmetries demonstrate deep
connections to topological physics such as topological insulators\citep{Weng_Fang_2014_4_11002__Transition},
topological crystalline insulators\citep{Fu_Fu_2011_106_106802__Topological,Hsieh_Fu_2012_3_982__Topological},
Dirac and Weyl semimetals\citep{Wang_Fang_2012_85_195320__Dirac,Weng_Dai_2015_5_11029__Weyl},
nodal line and nodal loop semimetals\citep{Fang_Fu_2015_92_81201__Topological,Li_Yang_2017_96_81106__Type,Ma_Yao_2018_98_201104__Mirror},
nodal chain metals\citep{Bzdusek_Soluyanov_2016_538_75__Nodal}, hourglass-band
materials \citep{Wang_Bernevig_2016_532_189__Hourglass,Li_Yang_2018_97_45131__Nonsymmorphic,Fu_Yao_2018_98_75146__Hourglasslike},
topological photonic crystals with all-dielectric materials\citep{Wu_Hu_2015_114_223901__Scheme,Lu_Soljacic_2016_12_337__Symmetry,Slobozhanyuk_Khanikaev_2016_11_130__Three,Ji_Yao_2019_99_43801__Transport}.
In addition, representation theory of space group is also used in
symmetry indicator method or elementary band representation method
to classify symmetry-protected band topology of nonmagnetic materials\citep{Po_Watanabe_2017_8_50__Symmetry,Kruthoff_Slager_2017_7_41069__Topological,Song_Fang_2018_9_3530_1711.11049v3_Quantitative,Zhang_Fang_2019_566_475__Catalogue,Tang_Wan_2019_15_470__Efficient,Tang_Wan_2019_566_486__Comprehensive,Tang_Wan_2019_5_8725__Topological,Bradlyn_Bernevig_2017_547_298__Topological,Cano_Bernevig_2018_120_266401__Topology,Cano_Bernevig_2018_97_35139__Building,Vergniory_Wang_2019_566_480__complete},
which greatly helps to search materials with specific band topology
systematically.

Although nearly all modern codes of density functional theory (DFT)
support symmetry analysis such as doing the Brillouin zone (BZ) integration
in a symmetrically irreducible wedge region of BZ to accelerate calculation,
none of them gives full information of space groups and their representations
as far as we know. DFT codes \textsf{VASP}\citep{vasp1} and \textsf{ABINIT}\citep{abinit1}
gives SG operations in their output files and \textsf{ABINIT} also
provides the SG name, but neither of them give the information about
the little-group (LG) irreducible representations (IRs). It's known
that SG IR is obtained by inducing from LG IR (also called small representation),
and in most cases LG IR is enough for symmetry analysis. DFT codes
\textsf{WIEN2k}\citep{wien2k1} and \textsf{Quantum ESPRESSO}\citep{QuantumESPRESSO1}
can give LG IRs in the simple case of symmorphic space groups, but
neither of them can process LG IRs with wave vectors (i.e. k-points)
on the boundary of BZ for nonsymmorphic space groups. The manual of
\textsf{WIEN2k} says \textquotedblleft \textit{It will not work in
cases of non-symmorphic spacegroups AND k-points at the surface of
the BZ}\textquotedblright{} for its program \textsf{IRREP}, while
in the output of \textsf{band.x} of \textsf{Quantum ESPRESSO} there
are remarks ``\textit{zone border point and non-symmorphic group,
symmetry decomposition not available}''.

In order to determine the LG IRs of Bloch states there are mainly
two steps: the first step is to calculate the characters of each operation
in the little group, and the second step is to identify the LG IRs
by looking up the character tables of LG IRs. In fact, the character
tables of LG IRs were given by several books decades ago\citep{Kovalev1965,MLbook,ZakCGG,BCbook,CDML}.
Then why few DFT codes can give LG IRs in all cases? We think the
main reason is that the large amount of data relating to all LG IRs
and the not-so-easy relations among the data to those who are not
very familiar with the representation theory of space groups hinder
the realization of full support of LG IRs in the DFT codes. On the
contrary, in the simple case of symmorphic space groups which can
be treated by \textsf{WIEN2k} and \textsf{Quantum ESPRESSO}, only
PG character tables are needed to determine LG IRs, and PG character
tables have much smaller data amount of several pages compared to
those of LG IRs which can fill a book of hundreds of pages. 

In fact, there have been third-party databases of SG/LG IRs available
for a long time, i.e. the \textsf{ISOTROPY} software suite\citep{iso1,iso-ir}
and the Bilbao Crystallographic Server (BCS)\citep{BCS-II,BCS-DSG}.
The \textsf{ISOTROPY} is a collection of programs using space group
theory to analyze phase transitions in crystals. It contains and provides
all the data of SG IRs, but it does not give the data of LG IRs directly.
And it does not have IR data of double space groups either, because
they are not needed to analyze phase transitions in crystals. On the
other hand, the BCS is a user-friendly website with various crystallographic
databases and programs available online. It has full support of LG
IRs and SG IRs for both space groups and double space groups. However,
all the IR data are online and BCS does not provide offline programs.
For common users, this makes BCS not convenient for batch processing
a large number of offline jobs. Only recently, has there been a program
called \textsf{irvsp} capable of calculating the LG IRs of Bloch states\citep{Gao_Wang_2020___2002.04032v1_Irvsp}.
The \textsf{irvsp} is a post-processing program written in fortran
and designed to calculate the LG IRs of Bloch states generated by
\textsf{VASP} according to the character tables of LG IRs of BCS (obtained
from a developer of BCS).

\begin{figure}
\begin{centering}
\includegraphics[width=10cm]{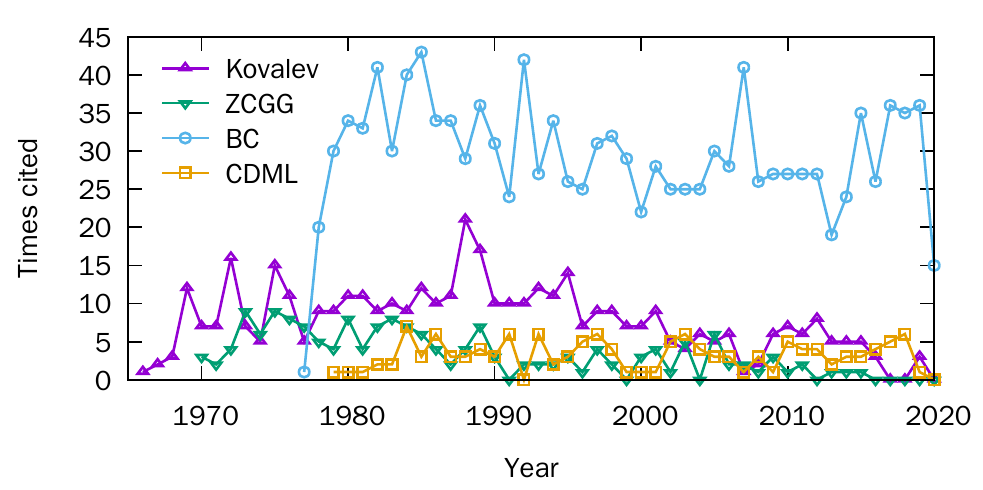}
\par\end{centering}
\caption{Times cited of the books \citep{Kovalev1965}, \citep{ZakCGG}, \citep{BCbook},
and \citep{CDML} in each year on Web of Science, corresponding to
the Kovalev, ZCGG, BC, and CDML conventions of space groups and their
IRs respectively. The data are up to Aug. 2020.\label{fig:timesCited}}
\end{figure}

Historically, there are different conventions used to describe space
groups and their IRs by different authors, such as Kovalev convention\citep{Kovalev1965},
Zak-Casher-Gl\"{u}ck-Gur (ZCGG) convention\citep{ZakCGG}, Bradley-Cracknell
(BC) covention\citep{BCbook}, and Cracknell-Davies-Miller-Love (CDML)
convention\citep{CDML}. Both \textsf{ISOTROPY} and BCS use the CDML
convention. However, after we analyzed the times cited of these conventions
on Web of Science\citep{note1} , we found that the BC convention
is the most used one, especially in recent years, as shown in Fig.
\ref{fig:timesCited}. This is not surprising, because the BC book
not only contains the tables of LG IRs and related complete data but
also contains comprehensive and systematic theory of space group and
its representation, which makes it a classic reference book and teaching
material. In addition, the BC book \citep{BCbook} is on sale and
hence the most easily obtained one while the other books \citep{Kovalev1965,ZakCGG,CDML}
are all out of print and hardly obtained. For example, we have tried
our best to seek the CDML book \citep{CDML} and failed finally, and
hence we can only understand the CDML convention from \textsf{ISOTROPY}
or BCS indirectly. The first version of the BC book was published
in 1972 and a reprint version was published in 2009. We think, it
was the popularity of the BC book that led to its reprint in 2009,
and the reprint made the BC book easily obtained and more popular. 

Although the BC book is popular, the space group settings used in
it are different from the commonly used settings in ``International
Tables for Crystallography, Volume A'' (hereafter referred to as
ITA)\citep{ITA}. This maybe make the BC settings of space groups
not very intuitive. Additionally, there are various tables correlated
with each other in the BC book, which makes it somewhat tedious and
complicated to extract information from these tables. For example,
if we want to know the character of the operation $\{\sigma_{z}|\frac{1}{4}\frac{1}{4}\frac{1}{4}\}$
in the IR $\Gamma_{2}^{+}$ of space group $Fddd$ (No. 70) we have
to first look up the Tab. 5.7 in the BC book (hereafter referred to
as ``BC-Tab. 5.7'') to find the abstract group $G_{8}^{3}$, the
generators of the Herring little group (HLG) $\{C_{2z}|000\}$, $\{C_{2y}|000\}$,
$\{I|\frac{1}{4}\frac{1}{4}\frac{1}{4}\}$, and the letter ``b''
according to which we can know that the IR $\Gamma_{2}^{+}$ is the
IR $R_{2}$ of $G_{8}^{3}$ from BC-Tab. 5.8. Then we calculate the
elements of HLG according to the generators $P=\{C_{2z}|000\}$, $Q=\{C_{2y}|000\}$,
$R=\{I|\frac{1}{4}\frac{1}{4}\frac{1}{4}\}$ of $G_{8}^{3}$ in BC-Tab.
5.1 and find that $C_{6}=PR=\{\sigma_{z}|\frac{1}{4}\frac{1}{4}\bar{\frac{3}{4}}\}$.
Hence we find in the character table of BC-Tab. 5.1 that the character
of $\{\sigma_{z}|\frac{1}{4}\frac{1}{4}\bar{\frac{3}{4}}\}$ in IR
$R_{2}$ of $G_{8}^{3}$ is $-1$. According to the properties of
LG IR {[}refer to Eq. (\ref{eq:Gammakpv+t}){]} we know that the character
of $\{\sigma_{z}|\frac{1}{4}\frac{1}{4}\frac{1}{4}\}$ in IR $\Gamma_{2}^{+}$
is $\chi(\{\sigma_{z}|\frac{1}{4}\frac{1}{4}\frac{1}{4}\})=e^{-i\vk\cdot\Delta\vR}\chi(\{\sigma_{z}|\frac{1}{4}\frac{1}{4}\bar{\frac{3}{4}}\})=\chi(\{\sigma_{z}|\frac{1}{4}\frac{1}{4}\bar{\frac{3}{4}}\})=-1$,
in which $\vk$ is $\Gamma$ and $\Delta\vR=(\frac{1}{4}\frac{1}{4}\frac{1}{4})-(\frac{1}{4}\frac{1}{4}\bar{\frac{3}{4}})=(001)$.
Furthermore, the rotation matrices used to calculate the HLG are defined
in BC-Tab. 3.2. This example shows the cumbersome process of extracting
information from the tables of the BC book. If lots of data are obtained
this way manually, it's not only tedious but also prone to error.
Consequently, a program based on the SG and IR data in the BC book
and capable of automating this process is highly required. However,
there are no such programs available as we know, therefore we developed
such a program package named \textsf{SpaceGroupIrep} in the Mathematica
language.

Different conventions use different notations to label the LG IRs,
therefore the meanings of IR labels are clear only if the convention
used is pointed out. It is particularly true for the ZCGG, BC, and
CDML conventions, because they use similar labels such as $\Gamma_{1},\Gamma_{2},\cdots,X_{1},X_{2},\cdots$
(called ``$\Gamma$ labels'' here) but with probably different meanings.
A concomitant problem is how to find the correspondence between the
IR labels of two different conventions. The only route we know before
our \textsf{SpaceGroupIrep} is to use \textsf{ISOTROPY}. \textsf{ISOTROPY}
can give the correspondence of IR labels for all the Kovalev, ZCGG,
BC, and CDML conventions. However, \textsf{ISOTROPY} works only for
high-symmetry (HS) k-points but not for HS lines, and \textsf{ISOTROPY}
does not distinguish a couple of complex conjugate IRs related by
time reversal symmetry. Furthermore, \textsf{ISOTROPY} does not support
IRs of double space groups. Based on these reasons, at present we
have realized the correspondence of LG IR labels between BC convention
and CDML convention in \textsf{SpaceGroupIrep}. Notice that hereafter
the CDML convention will be called BCS convention, because the CDML
IR data we used are actually the BCS IR data collected from the output
of \textsf{irvsp}.

The \textsf{SpaceGroupIrep} package contains all necessary data related
to space groups and their IRs defined in the BC book and tens of functions
manipulating the data. It can give both the LG IRs and SG IRs at any
k-point in an intuitive table form. It can calculate the reduction
of the direct product of two SG IRs. It can read the \textsf{trace.txt}
file generated by \textsf{vasp2trace}\citep{Vergniory_Wang_2019_566_480__complete}
and determine the LG IRs in BC convention for all Bloch states. It
can give the correspondence of LG IR labels between BC convention
and BCS convention. It can also convert any given crystalline structure
to the one in BC convention with the help of an external  package
\textsf{spglib}\citep{spglib}. The above aspects are also true for
double-valued IRs. In addition, \textsf{SpaceGroupIrep} can help study
and understand the BC book. It can easily give the elements of a designated
space group, little group, Herring little group, or central extension
of little co-group and calculate the multiplication of the elements.
In a word, the \textsf{SpaceGroupIrep} package is a database and tool
set for SG/LG IRs in BC convention, which is very useful in both study
and research. 

\section{Theory}

\subsection{Representation theory overview}

Let $G$ be a space group whose elements are in the form of Seitz
symbol $\{R|\bm{v}\}$. $\{R|\bm{v}\}$ means a rotation $R$ followed
by a translation by vector $\bm{v}$. Select one k-point from each
wave vector star (i.e. k-star) arbitrarily. Then the induced representations
of all the allowed LG IRs of these selected k-points are just all
the SG IRs of $G$. Let $\vk$ be a wave vector, its little group
be $G^{\vk}$, and its wave vector star be $^{*}\vk$. Suppose that
$\Gamma_{p}^{\vk}$ is the $p$-th allowed LG IR of $G^{\vk}$ with
dimension $d_{p}$. The modifier ``allowed'' means that $\Gamma_{p}^{\vk}$
satisfies 
\begin{equation}
\Gamma_{p}^{\vk}(\{E|\bm{t}\})=e^{-i\vk\cdot\bm{t}}\Gamma_{p}^{\vk}(\{E|\bm{0}\})=e^{-i\vk\cdot\bm{t}}I_{p},
\end{equation}
where $E$ is the identity element of point group, $\bm{t}$ is a
lattice vector, hence $\{E|\bm{t}\}$ is a pure translation operation,
and $I_{p}$ is a $d_{p}\times d_{p}$ identity matrix. This allowing
condition makes LG IRs compatible with the IRs of translation group
$T$, and it also makes the representation matrices of all LG elements
with the same rotation easily obtained through the relation
\begin{equation}
\Gamma_{p}^{\vk}(\{R|\bm{v}+\bm{t}\})=e^{-i\vk\cdot\bm{t}}\Gamma_{p}^{\vk}(\{R|\bm{v}\})\label{eq:Gammakpv+t}
\end{equation}
if $\Gamma_{p}^{\vk}(\{R|\bm{v}\})$ is known. If not explicitly stated
otherwise, all the LG IRs we mentioned are allowed. Use $\{R_{\alpha}|\bm{\tau}_{\alpha}\}$
$(\alpha=1,2,\cdots,m_{\vk})$ to denote the coset representatives
of the left cosets of $G^{\vk}$ in $G$. Then the SG IR induced from
$\Gamma_{p}^{\vk}$, denoted by $\Gamma_{p}^{\vk}\uparrow G$ or $^{*}\Gamma_{p}^{\vk}$
, is an $m_{\vk}d_{p}$-dimensional IR and is determined by
\begin{equation}
\big[{}^{*}\Gamma_{p}^{\vk}(\{R|\bm{v}\})\big]_{\alpha\beta}=\begin{cases}
\Gamma_{p}^{\vk}(\{R_{\alpha}|\bm{\tau}_{\alpha}\}^{-1}\{R|\bm{v}\}\{R_{\beta}|\bm{\tau}_{\beta}\}) & \ \ \text{if }\{R_{\alpha}|\bm{\tau}_{\alpha}\}^{-1}\{R|\bm{v}\}\{R_{\beta}|\bm{\tau}_{\beta}\}\in G^{\vk}\\
0 & \ \ \text{otherwise}
\end{cases},
\end{equation}
where $m_{\vk}=|G|/|G^{\vk}|$ is the number of k-points in the star
$^{*}\vk$, and $|G|$ means the order of the group $G$. It can be
seen that LG IRs have to be known first to determine SG IRs. Consequently,
the core problem in SG representation theory is to obtain all LG IRs
of a space group.

The projective representation method is a general method to obtain
LG IRs. Suppose $\{R_{i}|\bm{v}_{i}\}$ is one of the coset representatives
of the cosets of $T$ in $G^{\vk}$, and then the set of $R_{i}$
is the little co-group of $G^{\vk}$, denoted by $\bar{G}^{\vk}.$
Further suppose that $\tilde{D}_{p}^{\vk}$ is the $p$-th projective
representation of $\bar{G}^{\vk}$ with factor system $\mu(R_{i},R_{j})=\exp[-i(R_{i}^{-1}\vk-\vk)\cdot\bm{v}_{j}]$,
then the LG IR $\Gamma_{p}^{\vk}$ is determined by the following
simple relation
\begin{equation}
\Gamma_{p}^{\vk}(\{R|\bm{v}\})=e^{-i\vk\cdot\bm{v}}\tilde{D}_{p}^{\vk}(R).
\end{equation}
To obtain the projective representation $\tilde{D}_{p}^{\vk}$ of
the little co-group $\bar{G}^{\vk}$, we can resort to the central
extension of $\bar{G}^{\vk}$, denoted by $\bar{G}^{\vk*}$. The group
elements of $\bar{G}^{\vk*}$ are in the form of $(R_{i},\alpha)$
with $\alpha=0,1,\cdots,g-1$ and $g$ is the smallest positive integer
determined by the factor system
\begin{equation}
\mu(R_{i},R_{j})=\exp[-i(R_{i}^{-1}\vk-\vk)\cdot\bm{v}_{j}]=\exp[2\pi ia(R_{i},R_{j})/g]\label{eq:facSys}
\end{equation}
for all $i,j,$ in which the function $a(R_{i},R_{j})$ determined
by $\mu(R_{i},R_{j})$ has integer value also in range $[0,g-1]$.
The group multiplication of central extension $\bar{G}^{\vk*}$ is
defined as 
\begin{equation}
(R_{i},\alpha)(R_{j},\beta)=(R_{i}R_{j},\;\alpha+\beta+a(R_{i},R_{j})\!\!\mod\,g)\label{eq:CEtimes}
\end{equation}
with the property 
\begin{equation}
(R_{i},\alpha)=(R_{i},0)(E,\alpha)=(E,\alpha)(R_{i},0).
\end{equation}
Then all the irreducible projective representations we need can be
obtained from the allowed ordinary IRs of the corresponding central
extension. Suppose $\Delta_{p}^{\vk}$ is the $p$-th allowd IR of
$\bar{G}^{\vk*}$ with the property
\begin{equation}
\Delta_{p}^{\vk}((E,\alpha))=e^{i2\pi\alpha/g}I_{p},
\end{equation}
and then the irreducible projective representation $\tilde{D}_{p}^{\vk}$
is determined by
\begin{equation}
\tilde{D}_{p}^{\vk}(R_{i})=\Delta_{p}^{\vk}((R_{i},0)).
\end{equation}

Apart from the projective representation method which is available
for any k-point, there is a Herring little group method which is easier
but only available for HS k-points. The HLG of $\vk,$ denoted by
$\herring{G^{\vk}}$, is a quotient group defined by $\herring{G^{\vk}=G^{\vk}/T^{\vk}}$
in which $T^{\vk}$ is a subgroup of $T$ with all translations $\{E|\bm{t}\}$
satisfying $e^{-i\vk\cdot\bm{t}}$=1. When $\vk$ is a HS k-point
, $T^{\vk}$ is an infinite group and hence $\herring{G^{\vk}}$ is
a finite group whose order is not very large. Then the IRs of $G^{\vk}$
can be obtained directly from the IRs of $\herring{G^{\vk}}$. Suppose
$\herring{D_{p}^{\vk}}$ is the $p$-th IR of $\herring{G^{\vk}}$
and then the LG IR $\Gamma_{p}^{\vk}$ of $G^{\vk}$ is determined
by
\begin{equation}
\Gamma_{p}^{\vk}(\{R_{i}|\bm{v}_{i}\})=\herring{D_{p}^{\vk}(}\{R_{i}|\bm{v}_{i}\}T^{\vk}).
\end{equation}
When $\vk$ is not a HS k-point, $\herring{G^{\vk}}$ is generally
an infinite group whose IRs are not easily obtained. In this case
the HLG method loses its advantage over the projective representation
method. Accordingly, the BC book uses both the two methods to describe
the LG IRs in BC-Tabs. 5.7 and  6.13, i.e., it uses HLG to describe
HS k-points and uses central extension of little co-group to describe
k-points on HS lines. And each HLG or central extension is isomorphic
to a certain abstract group $G_{m}^{n}$ whose IRs are known and given
in BC-Tab. 5.1.

\subsection{Brillouin zone and k-points}

\begin{figure}
\begin{centering}
\includegraphics[width=15cm]{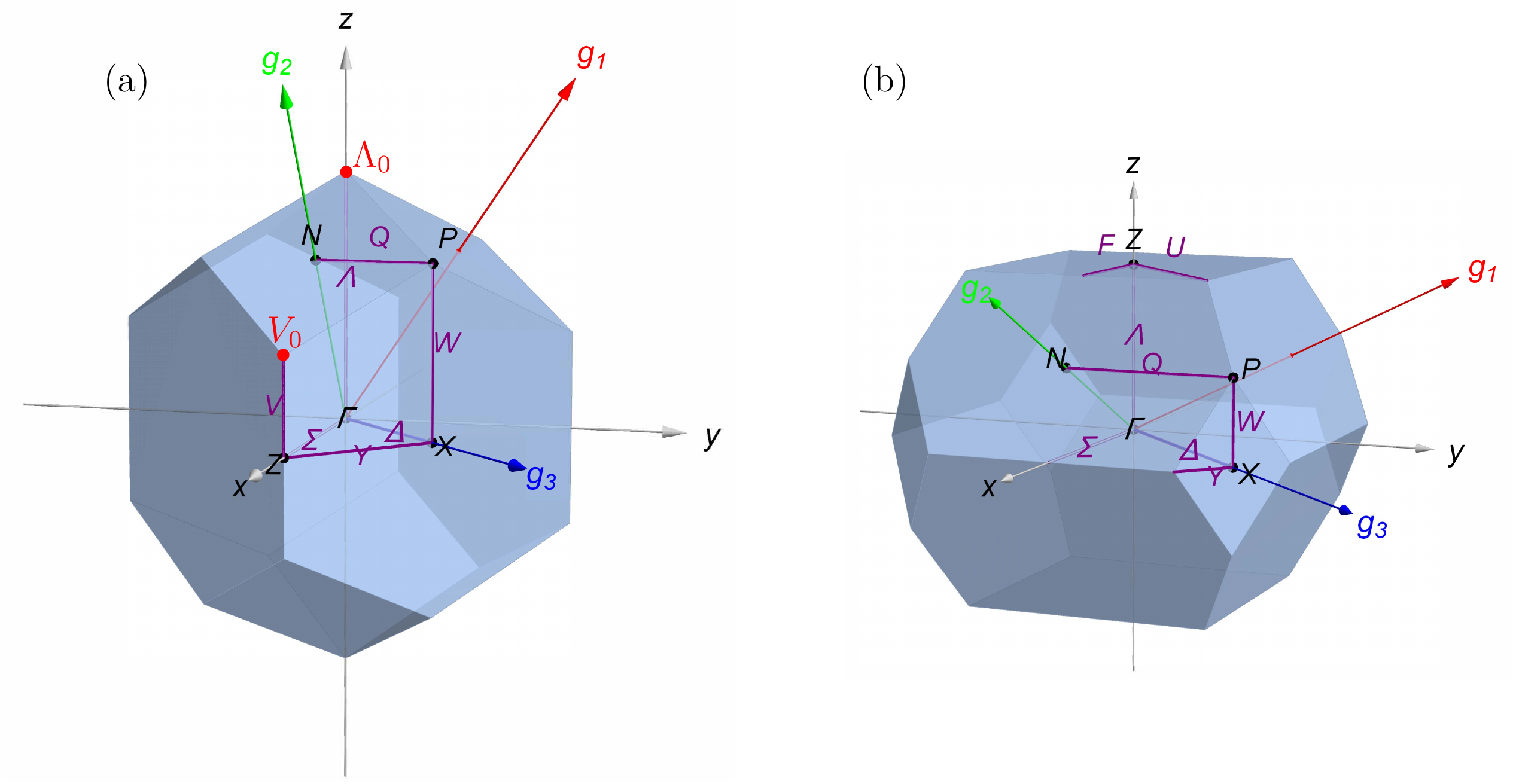}
\par\end{centering}
\caption{Two different types of BZs for body-centered tetragonal Bravais lattice
with lattice constants $a$ and $c$: (a) for $a>c$ and (b) for $a<c$.
Black capital letters are HS k-points, and purple ones are HS lines.
Except the two auxiliary points $\Lambda_{0}$ and $V_{0}$, the two
BZs are generated by \lstinline!showBZDemo["TetrBody(a)"]! and \lstinline!showBZDemo["TetrBody(b)"]!
respectively.\label{fig:tetrBZ}}
\end{figure}

Generally, the Wigner-Seitz unit cell in reciprocal space is used
as the (first) BZ. But for triclinic and monotonic Bravais lattices,
their Wigner-Seitz BZs are dependent very much on the actual values
of the lattice parameters and are difficult to draw or visualize.
Therefore, in practice, the BC book uses the reciprocal primitive
cell, i.e. a parallelepiped centered at $\vk=0$, as the BZ for triclinic
and monotonic Bravais lattices (and so does BCS). For other Bravais
lattices, Wigner-Seitz BZs are used in the BC book. In spite of defining
BZ this way, the shape of BZ is not always unique to each Bravais
lattice. Depending on the ratios of lattice constants, there are more
than one type (or shape) of BZ for base-centered orthorhombic {[}(a),
(b) two types{]}, body-centered orthorhombic {[}(a), (b), (c) three
types{]}, face-centered orthorhombic {[}(a), (b), (c), (d) four types{]},
body-centered tetragonal {[}(a), (b) two types, see Fig. \ref{fig:tetrBZ}{]},
and trigonal {[}(a), (b) two types{]} Bravais lattices. These five
kinds of Bravais lattices are called ``multiple-BZ Bravais lattices''.
There are in total 22 different types of BZs for the 14 Bravais lattices
which are shown in BC-Figs. 3.2 to 3.15.

HS k-points and k-points on HS lines in the basic domain are defined
in BC-Tab. 3.6, and all the k-points in BC-Tab. 3.6 are termed ``BC
standard k-points'' here. For multiple-BZ Bravais lattices, one k-point
name may have different coordinates for different BZ types, but these
coordinates are equivalent (differing by a reciprocal lattice vector)
to each other with only one exception which is the $F$ k-point of
trigonal lattice with coordinates $(0\frac{1}{2}\bar{\frac{1}{2}})$
and $(\frac{1}{2}\frac{1}{2}0)$ for BZ types (a) and (b) respectively.
This means that for a space group the LG IRs for a given k-point name
do not depend on BZ types, except for $F$ k-point in space groups
of trigonal lattice. This is demonstrated as the two entries ``(a)
$F$'' and ``(b) $F$'' existing in BC-Tabs. 5.7 and  6.13 for
space groups of trigonal lattice. 

Also for a multiple-BZ Bravais lattice, k-points on some HS lines
are named only for certain BZ types but not for others. Take body-centered
tetragonal lattice for example, $V$ HS line only exists in (a) type
BZ, while $F$ and $U$ HS lines only exist in (b) type BZ, as shown
in Fig. \ref{fig:tetrBZ}. In fact, both $\Lambda$ and $V$ in Fig.
\ref{fig:tetrBZ}(a) correspond to the $\Lambda$ in Fig. \ref{fig:tetrBZ}(b).
This can be analyzed as follow. The $\Lambda$ HS lines in both (a)
and (b) type BZs have coordinates $(uu\bar{u})$ (here we use $u$
for the $\alpha$ in BC-Tab. 3.6), but the range of $u$ is different.
The $\Lambda$ in (b) type BZ is from $\Gamma$ $(000)$ to $Z$ $(\frac{1}{2}\frac{1}{2}\bar{\frac{1}{2}})$
with $u\in(0,\frac{1}{2})$, while the $\Lambda$ in (a) type BZ is
from $\Gamma$ $(000)$ to $\Lambda_{0}$ $(\lambda_{0}\lambda_{0}\bar{\lambda}_{0})$
with $u\in(0,\lambda_{0}]$ and $\lambda_{0}=\frac{1}{4}+\frac{c^{2}}{4a^{2}}<\frac{1}{2}$
($a$ and $c$ are lattice constants and $a>c$ for (a) type BZ).
The $V$ in (a) type BZ with coordinates $(-\frac{1}{2}+u,\frac{1}{2}+u,\frac{1}{2}-u)$
is from $Z$ $(u=0)$ to $V_{0}$ $(u=v_{0})$ and $v_{0}=\frac{1}{4}-\frac{c^{2}}{4a^{2}}$.
We can see that $\lambda_{0}+v_{0}=\frac{1}{2}$. So, if $\lambda_{0}<u<\frac{1}{2}$,
the k-point $(uu\bar{u})$ lies on the extension line of $\Lambda$
outside the BZ of type (a). However, if this point $(uu\bar{u})$
is translated by $-\bm{g}_{2}$ to $(u,u-1,-u)$ and then transformed
by inversion $I$ to $(-u,1-u,u)$, it just lies on the line segment
$V$. This becomes clear if we do a substitution $u=\frac{1}{2}-u'$
and $(-u,1-u,u)$ becomes $(-\frac{1}{2}+u',\frac{1}{2}+u',\frac{1}{2}-u')$
with $0<u'<\frac{1}{2}-\lambda_{0}=v_{0}$. 

\subsection{LG IRs at any k-point\label{subsec:LGIR-at-any-k}}

Note that the LG IRs given in BC-Tabs. 5.7 and 6.13 are only directly
for BC standard k-points, i.e. those defined in BC-Tab. 3.6, not for
every k-point. Fortunately, LG IRs at any k-point can be obtained
from the LG IRs in BC-Tabs. 5.7 and 6.13 according to certain transformation
relations, except the k-points on $Z'$ HS line for space group $Pa\bar{3}$
(No. 205) whose LG IRs have to be given additionally in BC-Tabs. 5.11
and 6.15. Therefore, the $Z'$ with coordinates $(\frac{1}{2}u0)$
has to be added to the BC standard k-points for space group No. 205,
and complete BC LG IR tables comprise BC-Tabs. 5.7, 5.11, 6.13, and
6.15. To describe LG IRs at any k-point, the problem of naming k-point
has to be solved firstly. Any k-point $\vk$ can be classified as
one of the five types as follow.
\begin{itemize}
\item Type I, k-point which is identical to the BC standard k-point $\vk_{{\rm BC}}$
or equivalent to $\vk_{{\rm BC}}$, i.e. $\vk\equiv\vk_{{\rm BC}}$
($\equiv$ means the equivalence of k-points).
\item Type II, k-point not equivalent to $\vk_{{\rm BC}}$ but equivalent
to one arm of $^{*}\vk_{{\rm BC}}$ (the star of $\vk_{{\rm BC}}$),
which means that there is an element $\{S|\bm{w}\}\in G$ such that
$\vk\equiv S\vk_{{\rm BC}}$.
\item Type III, k-point not equivalent to any arm of $^{*}\vk_{{\rm BC}}$.
But there is an element $\{S|\bm{w}\}\notin G$ satisfying $\{S|\bm{w}\}^{-1}G\{S|\bm{w}\}=G$
such that $\vk\equiv S\vk_{{\rm BC}}.$
\item Type IV, general k-point whose little co-group has only identity element
and which does not belong to types I\textendash III.
\item Type V, k-point not belonging to types I\textendash IV. In fact, this
$\vk$ is either on a HS plane with little co-group $\{E,\sigma\}$
($\sigma$ is mirror reflection) for space groups other than $P\bar{1}$
(No. 2), or a HS k-point with little co-group $\{E,I\}$ for space
group $P\bar{1}$. 
\end{itemize}
The names of type I k-points are directly defined in the BC book.
For k-points of type II and III, we usually borrow the name of $\vk_{{\rm BC}}$
to name $\vk$ if $\vk\equiv S\vk_{{\rm BC}}$. But it should be kept
in mind that this is only an expedient and when necessary a name different
from $\vk_{{\rm BC}}$ has to be used for $\vk$ to avoid confusion.
The type IV k-point is simply named ``GP''. Note that not all general
k-points are named ``GP'' because some are BC standard k-points
which have been named, e.g. all the k-points with the abstract group
$G_{1}^{1}$ in BC-Tab. 5.7. Type V k-points comprise all k-points
on HS planes and some HS k-points of space group $P\bar{1}$, whose
names are not defined in the BC book. Accordingly for simplicity,
we just use ``UN'' as the name of these unnamed k-points of type
V. When necessary, customized names can be used to replace ``UN''.

The LG IRs of type I k-point $\vk_{{\rm BC}}$ are given in the BC
LG IR tables. For a k-point $\vk$ of type II and III, its little
group $G^{\vk}$ is isomorphic to $G^{\vk_{{\rm BC}}}$ because they
are conjugate to each other, i.e. $\{S|\bm{w}\}G^{\vk_{{\rm BC}}}\{S|\bm{w}\}^{-1}=G^{\vk}$.
Therefore, the LG IRs of $G^{\vk}$ can be obtained from those of
$G^{\vk_{{\rm BC}}}$ by 
\begin{equation}
\Gamma_{p}^{\vk}(\{R|\bm{v}\})=\Gamma_{p}^{\vk_{{\rm BC}}}(\{S|\bm{w}\}^{-1}\{R|\bm{v}\}\{S|\bm{w}\})\ \ \ \text{for\ \ \ensuremath{\forall\{R|\bm{v}\}\in G^{\vk}}.}\label{eq:GMkGMkBC}
\end{equation}
The LG IRs for a k-point of type IV and V are not given in the BC
book and have to be calculated by ourselves. However, the calculations
are easy because of the low symmetry of the k-point. The central extension
of a GP k-point is trivially $G_{1}^{1}$ ($G_{2}^{1}$ for double-valued
IR), and the central extension of a UN k-point is either $G_{2}^{1}$
or $G_{4}^{1}$ ($G_{4}^{1}$, $G_{4}^{2}$, or $G_{8}^{2}$ for double-valued
IR). 

\section{Files and installation }

The Mathematica package \textsf{SpaceGroupIrep} mainly includes four
files: \textsf{SpaceGroupIrep.wl}, \textsf{AbstractGroupData.wl},
\textsf{LittleGroupIrepData.wl}, and \textsf{allBCSkLGdat.mx}. \textsf{SpaceGroupIrep.wl}
is the main file containing most functions and data and the other
three are all data files. \textsf{AbstractGroupData.wl} contains the
abstract group data in BC-Tab. 5.1 which are stored in \lstinline!AGClasses!,
\lstinline!AGCharTab!, and \lstinline!AGIrepGen!. \textsf{LittleGroupIrepData.wl
}contains the data of LG IRs in BC-Tabs. 5.7, 5.11, 6.13, and 6.15
which are stored in \lstinline!LGIrep! and \lstinline!DLGIrep! for
single-valued and double-valued representations respectively. \textsf{allBCSkLGdat.mx
}contains the BCS data of LG IRs collected from the output of \textsf{irvsp}.
To install the package \textsf{SpaceGroupIrep}, just create a directory
 \textsf{SpaceGroupIrep} containing the four files and place it under
any of the following paths:
\begin{itemize}
\item \lstinline!$InstallationDirectory/AddOns/Packages/!
\item \lstinline!$InstallationDirectory/AddOns/Applications/!
\item \lstinline!$BaseDirectory/Applications/!
\item \lstinline!$UserBaseDirectory/Applications/!
\end{itemize}
where \lstinline!$InstallationDirectory! is the installation directory
of Mathematica (version $\ge$ 11.2), and \lstinline!$BaseDirectory!
and \lstinline!$UserBaseDirectory! are the directories containing
respectively systemwide and user-specific files loaded by Mathematica.
The concrete values of \lstinline!$InstallationDirectory!, \lstinline!$BaseDirectory!,
and \lstinline!$UserBaseDirectory! can be obtained by running them
in Mathematica because they are all built-in symbols. Then one can
use the package after running \lstinline[literate={`}{\textasciigrave}{1}]!<<"SpaceGroupIrep`"!.

\section{Group elements and multiplication}

Following the notations in the BC book, we use $\bm{t}_{1},\bm{t}_{2},\bm{t}_{3}$
and $\bm{g}_{1},\bm{g}_{2},\bm{g}_{3}$ to represent the basic vectors
of the primitive cell and the reciprocal primitive cell respectively
which have the relations $\bm{t}_{i}\cdot\bm{g}_{j}=2\pi\delta_{ij}.$
$\bm{t}_{1},\bm{t}_{2},\bm{t}_{3}$ and $\bm{g}_{1},\bm{g}_{2},\bm{g}_{3}$
are defined in BC-Tab. 3.1 and BC-Tab. 3.3 respectively for each of
the 14 Bravais lattices. Then a real space vector $\bm{v}=v_{1}\bm{t}_{1}+v_{2}\bm{t}_{2}+v_{3}\bm{t}_{3}$
can be described by a column matrix of its coefficients (or coordinates)
$\mathsf{v}=(v_{1},v_{2},v_{3})^{T}$ with respect to $\bm{t}_{1},\bm{t}_{2},\bm{t}_{3}$;
similarly a wave vector $\vk=k_{1}\bm{g}_{1}+k_{2}\bm{g}_{2}+k_{3}\bm{g}_{3}$
can be described by $\mathsf{k}=(k_{1},k_{2},k_{3})^{T}$. Let $R$
be a rotation operation, and it rotates $\bm{v}$ to $\bm{v'}$ and
$\vk$ to $\vk'$, i.e. $\bm{v}'=R\bm{v}$ and $\vk'=R\vk$. If these
relations are described in matrix form they are $\mathsf{v'=R_{r}v}$
and $\mathsf{k'=R_{k}k}$, in which $\mathsf{R_{r}}$ and $\mathsf{R_{k}}$
are the rotation matrices of $R$ in real space and in reciprocal
space respectively. Note that both the coefficients of vectors and
the rotation matrices are dependent on basic vectors and hence on
the Bravais lattice, therefore for the same $R$ its $\mathsf{R_{r}}$
($\mathsf{R_{k}})$ is different for different Bravais lattices. $\mathsf{R_{r}}$
($\mathsf{R_{k}})$ is defined according to BC-Tab. 3.2 (BC-Tab. 3.4),
and for the same rotation $R$ there is relation 
\begin{equation}
\mathsf{R_{k}=}(\mathsf{R}_{r}^{-1})^{T}.
\end{equation}
In \textsf{SpaceGroupIrep}, we use functions \lstinline!getRotMat!
and \lstinline!getRotMatOfK! to get the rotation matrices $\mathsf{R_{r}}$
and $\mathsf{R_{k}}$ respectively according to their rotation names
(for available rotation names refer to BC-Tab. 3.2, and one prime
is replaced by one ``p'', e.g. \lstinline!"C21pp"! is used in the
code for $C_{21}''$), and inversely use \lstinline!getRotName! to
obtain the rotation name. All these functions use a string representing
Bravais lattice as their first argument which is listed in Tab. \ref{tab:brav}.
For example,

\begin{lstlisting}[backgroundcolor={\color{yellow!5!white}}]
|In[1]:=| getRotMat["OrthPrim","C2x"]
        getRotMat["OrthFace","C2x"]
        getRotMatOfK["OrthFace","C2x"]
        getRotName["OrthFace",{{0,0,1},{-1,-1,-1},{1,0,0}}]
|Out[1]=| {{-1,0,0},{0,1,0},{0,0,-1}}
|Out[2]=| {{0,0,1},{-1,-1,-1},{1,0,0}}
|Out[3]=| {{0,-1,1},{0,-1,0},{1,-1,0}}
|Out[4]=| C2x
\end{lstlisting}

\begin{table}
\caption{String codes representing Bravais lattices and available values for
\lstinline!BZtype! in package \textsf{SpaceGroupIrep}. The full string
code of a type of BZ is just its Bravais lattice string code for single-BZ
Bravais lattices, and is its Bravais lattice string code followed
by \lstinline!BZtype! in a pair of parentheses such as \lstinline!"OrthBase(a)"!
for multiple-BZ Bravais lattices. \label{tab:brav}}

\begin{centering}
\par\end{centering}
\begin{centering}
\begin{tabular}{llll}
\hline 
\multicolumn{2}{l}{~~~~~~~~Bravais lattice} & string code & \lstinline!BZtype!\tabularnewline
\hline 
Triclinic & primitive & \lstinline!"TricPrim"! & \lstinline!""!\tabularnewline
Monotonic & primitive & \lstinline!"MonoPrim"! & \lstinline!""!\tabularnewline
 & base-centered & \lstinline!"MonoBase"! & \lstinline!""!\tabularnewline
Orthorhombic & primitive & \lstinline!"OrthPrim"! & \lstinline!""!\tabularnewline
 & base-centered & \lstinline!"OrthBase"! & \lstinline!"a"!, \lstinline!"b"!\tabularnewline
 & body-centered & \lstinline!"OrthBody"! & \lstinline!"a"!, \lstinline!"b"!, \lstinline!"c"!\tabularnewline
 & face-centered & \lstinline!"OrthFace"! & \lstinline!"a"!, \lstinline!"b"!, \lstinline!"c"!, \lstinline!"d"!\tabularnewline
Tetragonal & primitive & \lstinline!"TetrPrim"! & \lstinline!""!\tabularnewline
 & body-centered & \lstinline!"TetrBody"! & \lstinline!"a"!, \lstinline!"b"!\tabularnewline
Trigonal & primitive & \lstinline!"TrigPrim"! & \lstinline!"a"!, \lstinline!"b"!\tabularnewline
Hexagonal & primitive & \lstinline!"HexaPrim"! & \lstinline!""!\tabularnewline
Cubic & primitive & \lstinline!"CubiPrim"! & \lstinline!""!\tabularnewline
 & face-centered & \lstinline!"CubiFace"! & \lstinline!""!\tabularnewline
 & body-centered & \lstinline!"CubiBody"! & \lstinline!""!\tabularnewline
\hline 
\end{tabular}
\par\end{centering}
\end{table}

For double space groups, we use \lstinline!{srot,o3det}! to describe
a spin rotation operation, where \lstinline!srot! is a SU(2) spin
rotation matrix defined in BC-Tab. 6.1 and \lstinline!o3det! is the
determinant (either 1 or $-1$) of corresponding O(3) rotation matrix.
Note that the SU(2) matrices in the BC book use \{spin down, spin
up\} as bases which has reversal sequence of the usually used \{spin
up, spin down\}. We use \lstinline!getSpinRotOp[rotname]! to get
the spin rotation operation according to the rotation name \lstinline!rotname!.
For rotations with an overbar such as $\bar{C}_{2z},$ their name
strings are all prefixed with \lstinline!bar! in the code, e.g. \lstinline!"barC2z"!
for $\bar{C}_{2z}.$ All available \lstinline!rotname!'s can be obtained
by \lstinline!Keys[getSpinRotOp]! because \lstinline!getSpinRotOp!
is in fact an association. Inversely, \lstinline!getSpinRotName[brav,{srot,o3det}]!
is used to obtain the rotation name, in which \lstinline!brav! is
the string code for Bravais lattice. In fact, here \lstinline!brav!
is only used to distinguish cubic compatible Bravais lattices (triclinic,
monoclinic, orthorhombic, tetragonal, and cubic) from hexagonal compatible
Bravais lattices (trigonal and hexagonal), because in each of these
two cases one rotation name is associated with only one SU(2) matrix.
For example,
\begin{lstlisting}[backgroundcolor={\color{yellow!5!white}}]
|In[1]:=| srop=getSpinRotOp["barC32+"]
        getSpinRotName["CubiBody",srop]
        getSpinRotName["CubiPrim",srop]
|Out[1]=| {{{-(1/2)-I/2,1/2-I/2},{-(1/2)-I/2,-(1/2)+I/2}},1}
|Out[2]=| barC32+
|Out[3]=| barC32+
\end{lstlisting}

Space group element $\{R|\bm{v}\}$ (or $\{R|v_{1}v_{2}v_{3}\}$)
is expressed as \lstinline!{R,{v1,v2,v3}}! in the \textsf{SpaceGroupIrep}
code where \lstinline!R! is the name string of the rotation $R$,
e.g. $\{C_{2z}|\frac{1}{2}\frac{1}{2}0\}$ expressed as \lstinline!{"C2z",{1/2,1/2,0}}!.
And the element $(R,\alpha)$ of the central extension is described
by \lstinline!{R,alpha}! in the code. We use \lstinline!getLGElem[sgno,k]!,
\lstinline!getHLGElem[sgno,kname]!, and \lstinline!getCentExt[sgno,kname]!
to get the elements of little group, HLG, and central extension respectively
for the space group of number \lstinline!sgno!, in which \lstinline!k!
can be either the k-point name or the k-point coordinates but \lstinline!kname!
can be only k-point name. These three functions also support double
space groups if an option \lstinline!"DSG"->True! is used. It should
be pointed out that the list of elements returned by \lstinline!getLGElem!
is actually the set of coset representatives of the cosets of $T$
in the little group of \lstinline!k!. In the same sense, the elements
of space group can be obtained by \lstinline!getLGElem! if \lstinline[mathescape=true]!k="$\Gamma$"!.
We also emphasize that although HLG is a quotient group whose elements
are cosets of $T^{\vk}$ in $G^{\vk}$, we usually omit $T^{\vk}$
and only care about the coset representatives which are also called
the elements of the HLG here for simplicity. Accordingly, \lstinline!getHLGElem!
returns the list of coset representatives of the real elements of
the HLG. Furthermore, for HLGs having the form of $G_{m}^{n}\otimes T_{q}$
in BC-Tabs. 5.7 and 6.13, only the abstract group $G_{m}^{n}$ is
needed to determine LG IRs, and hence the list of elements returned
by \lstinline!getHLGElem! is actually $G_{m}^{n}$ determined by
the generators given in BC-Tabs. 5.7 and  6.13. And the list returned
by \lstinline!getHLGElem! or \lstinline!getCentExt! has the same
element sequence as the corresponding abstract group in BC-Tab. 5.1.
Examples are as following:
\begin{lstlisting}[backgroundcolor={\color{yellow!5!white}},mathescape=true]
|In[1]:=| getLGElem[20,"$\Gamma$"]
        getLGElem[20,"R"]
        getHLGElem[20,"R"]
        getCentExt[20,"B"]
        getCentExt[20,"B","DSG"->True]
|Out[1]=| {{E,{0,0,0}}, {C2x,{0,0,0}}, {C2y,{0,0,1/2}}, {C2z,{0,0,1/2}}}
|Out[2]=| {{E,{0,0,0}}, {C2z,{0,0,1/2}}}
|Out[3]=| {{E,{0,0,0}}, {C2z,{0,0,1/2}}, {E,{0,0,1}}, {C2z,{0,0,3/2}}}
|Out[4]=| {{E,0}, {C2y,0}, {E,1}, {C2y,1}}
|Out[5]=| {{E,0}, {C2y,0}, {barE,1}, {barC2y,1},
         {barE,0}, {barC2y,0}, {E,1}, {C2y,1}}
\end{lstlisting}

The multiplication of two SG elements $\{R_{1}|\bm{v}\}\{R_{2}|\bm{w}\}=\{R_{1}R_{2}|R_{1}\bm{w}+\bm{v}\}$,
the inversion of a SG element \{$R|\bm{v}\}^{-1}=\{R^{-1}|-R^{-1}\bm{v}\},$
and the $n$-th power of a SG element $\{R|\bm{v}\}^{n}$ can be calculated
by \lstinline!SeitzTimes!, \lstinline!invSeitz!, and \lstinline!powerSeitz!
in the code respectively. All the three functions have double space
group versions with a \lstinline!DSG! prefix. Hence the functions
operating on SG elements are listed below.
\begin{lstlisting}[backgroundcolor={\color{yellow!5!white}}]
SeitzTimes[brav][{R1,{v1,v2,v3}},{R2,{w1,w2,w3}}]
invSeitz[brav][{R,{v1,v2,v3}}]
powerSeitz[brav][{R,{v1,v2,v3}},n]
DSGSeitzTimes[brav][{R1,{v1,v2,v3}},{R2,{w1,w2,w3}}]
DSGinvSeitz[brav][{R,{v1,v2,v3}}]
DSGpowerSeitz[brav][{R,{v1,v2,v3}},n]
\end{lstlisting}
In addition, multiplication of the central extension, i.e. Eq. (\ref{eq:CEtimes}),
is realized by the function \lstinline!CentExtTimes!, and its power
is realized by \lstinline!CentExtPower!. Also, the double space group
versions are available. They are listed below.
\begin{lstlisting}[backgroundcolor={\color{yellow!5!white}}]
CentExtTimes[brav,adict][{R1,alpha},{R2,beta}]
CentExtPower[brav,adict][{R,alpha},n]
DSGCentExtTimes[brav,adict][{R1,alpha},{R2,beta}]
DSGCentExtPower[brav,adict][{R,alpha},n]
\end{lstlisting}
In the above functions, \lstinline!adict! contains the information
of $a(R_{i},R_{j})$ defined in Eq. (\ref{eq:facSys}) and it can
be obtained by calling \lstinline!aCentExt[sgno,kname]! or \lstinline!aCentExt[brav,Gk,k]!
in which \lstinline!Gk! is the LG elements and \lstinline!k! is
the k-point coordinates. For double space group, the option \lstinline!"DSG"->True!
should be used in \lstinline!aCentExt!.

\section{Abstract group}

\begin{figure}
\begin{centering}
\includegraphics[width=9cm]{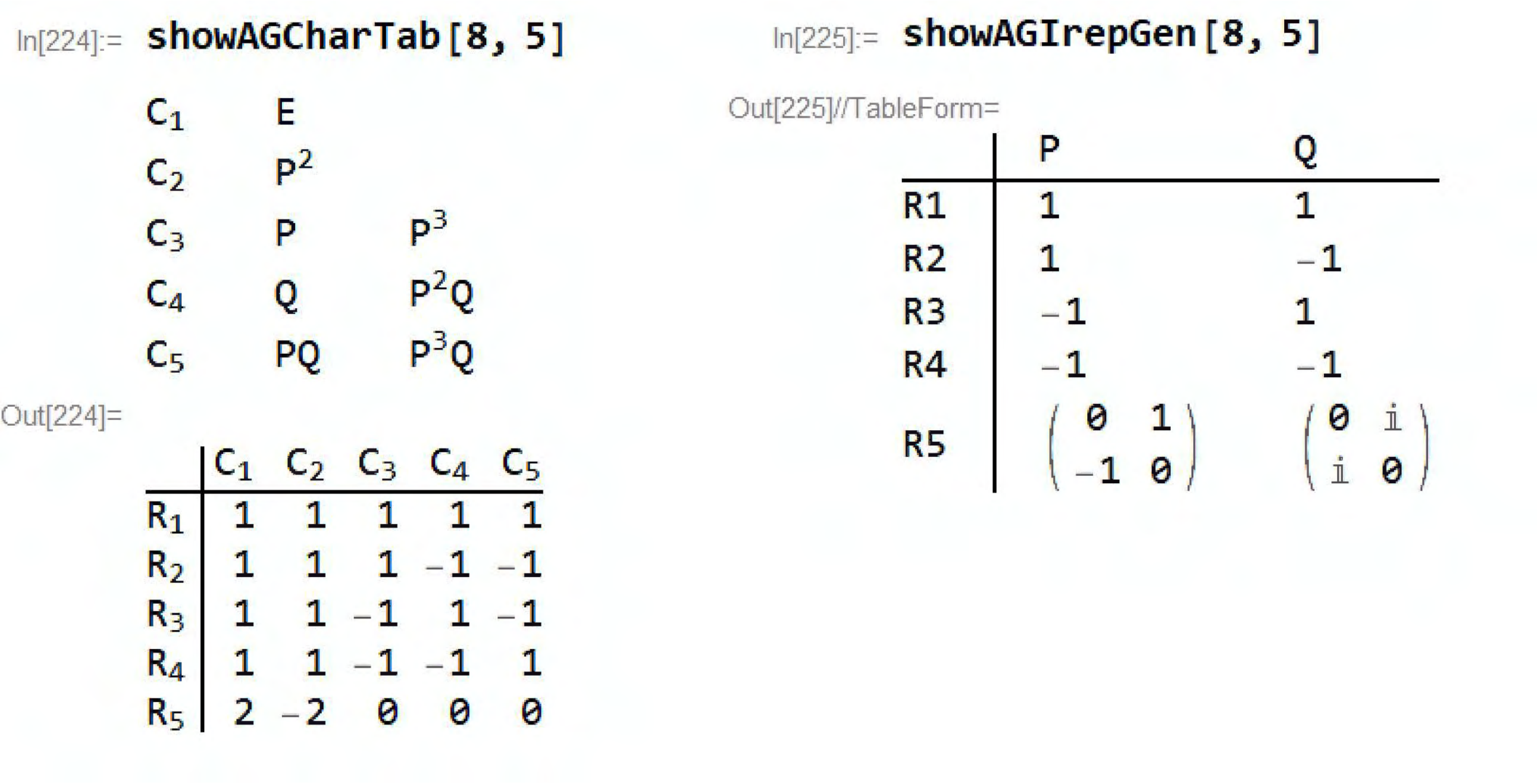}
\par\end{centering}
\caption{Examples of \lstinline!showAGCharTab[m,n]! and \lstinline!showAGIrepGen[m,n]!
for the abstract group $G_{8}^{5}$.\label{fig:showAGCharTab}}

\end{figure}

There are in total 93 abstract groups in BC-Tab. 5.1 which can be
used to describe LG IRs and whose indexes $\{m,n$\}'s are listed
in \lstinline!allAGindex!. The information of $G_{m}^{n}$ in BC-Tab.
5.1 is stored in \lstinline!AGClasses[m,n]!, \lstinline!AGCharTab[m,n]!,
and \lstinline!AGIrepGen[m,n]! which mean the classes, the character
table, and the generators of each IR respectively. The elements in
classes are described by their power exponents of the generators $P,Q,R,\cdots$,
e.g. the element $P^{2}QR^{3}$ is described by \lstinline!{2,1,3}!.
For ease of view, classes and character table are shown in table form
by \lstinline!showAGCharTab[m,n]!, and generators of each IR are
shown in table form by \lstinline!showAGIrepGen[m,n]!. Examples are
shown in Fig. \ref{fig:showAGCharTab}.

\section{Tables for LG IRs and SG IRs}

\subsection{Identify k-point}

When the coordinates of a k-point are given, we have to know its name
and its relation to the BC standard k-point before we can determine
its LG IRs. In other words, we have to classify the k-point into one
of the five types defined in subsection \ref{subsec:LGIR-at-any-k}
and find the operation $\{S|\bm{w}\}$ if the k-point is of type II
or III. There are two functions for doing this in \textsf{SpaceGroupIrep},
i.e. 
\begin{lstlisting}[backgroundcolor={\color{yellow!5!white}}]
identifyBCHSKpt[fullBZtype, kOrKlist]
identifyBCHSKptBySG[sgno, BZtypeOrBasVec, kOrKlist]  
\end{lstlisting}
in which \lstinline!fullBZtype! is the string code for one of the
22 types of BZs (see Tab. \ref{tab:brav}), \lstinline!kOrKlist!
is the numerical coordinates of a k-point or of a list of k-points,
and \lstinline!BZtypeOrBasVec! is either one of the \lstinline!BZtype!
in Tab. \ref{tab:brav} or the numerical basic vectors of the space
group. In fact, \lstinline!identifyBCHSKpt! is a preprocessor of
\lstinline!identifyBCHSKptBySG!. The former identifies a k-point
only according to BC-Tab. 3.6 without the SG information, while using
the SG information the latter can further determine the $\{S|\bm{w}\}$
based on the results of the former. 

The result returned by \lstinline!identifyBCHSKpt! is a list 
\begin{equation}
\{\vk,\text{\,kname},\,\text{\text{line\_info},\,}P(\vk),\,\vk_{{\rm BC}},\,S,\,S\vk_{{\rm BC}},\,\bm{g},\ {\rm u}\rightarrow v_{{\rm u}}\}\label{eq:kinfo0}
\end{equation}
where $\vk$ is the k-point to be identified, kname is the identified
name, line\_info is the connection of HS line (null string if $\vk$
is a HS point), $P(\vk)$ is the symmetry point group of $\vk$ in
the basic domain of the Bravais lattice, $\vk_{{\rm BC}}$ is the
BC standard k-point in the basic domain, $S$ is an element of $P(\vk)$
such that $\vk\equiv S\vk_{{\rm BC}}$, $\bm{g}=\vk-S\vk_{{\rm BC}}$
is the reciprocal lattice vector connecting $S\vk_{{\rm BC}}$ and
$\vk$, and ${\rm u}\rightarrow v_{{\rm u}}$ is the rule for the
actual value $v_{{\rm u}}$ of u. If kname is GP or UN only the first
four items exist in Eq. (\ref{eq:kinfo0}), and the last item ${\rm u}\rightarrow v_{{\rm u}}$
exists only when $\vk$ is on HS line. It is noteworthy that for multiple-BZ
Bravais lattice, a k-point can be identified as two points with different
names by \lstinline!identifyBCHSKpt!, i.e. two entries in the form
of Eq. (\ref{eq:kinfo0}) with different knames. And this case occurs
when the k-point is on certain HS lines. For example, the k-point
with coordinates $(-0.3,-0.3,0.3)$ is identified as $\Lambda$ in
the BZ of type \lstinline!"TetrBody(b)"! , while it is identified
as either $\Lambda$ or $V$ in the BZ of type \lstinline!"TetrBody(a)"!.
The code is as follow.

\begin{lstlisting}[backgroundcolor={\color{yellow!5!white}},mathescape=true]
|In[1]:=| identifyBCHSKpt["TetrBody(b)",{-0.3,-0.3,0.3}]
        identifyBCHSKpt["TetrBody(a)",{-0.3,-0.3,0.3}]
|Out[1]=| {{{-0.3,-0.3,0.3},$\Lambda$,$\Gamma$Z,C4v,{u,u,-u},C2x,{-u,-u,u},{0,0,0},u->0.3}}
|Out[2]=| {{{-0.3,-0.3,0.3},$\Lambda$,$\Gamma\Lambda$,C4v,{u,u,-u},C2x,{-u,-u,u},{0,0,0},u->0.3},
         {{-0.3,-0.3,0.3},V,ZV,C4v,{-1/2+u,1/2+u,1/2-u},E,
                                   {-1/2+u,1/2+u,1/2-u},{0,-1,0},u->0.2}}
\end{lstlisting}

\begin{table}[t]
\caption{The table of $u_{{\rm max}}$ for HS lines, i.e. the maximum value
of $u$ that keeps the BC standard k-point inside or on the boundary
of the BZ. Except the HS lines in this table, the $u_{{\rm max}}$
of all other HS lines in BC-Tab. 3.6 is $\frac{1}{2}$. $a,b,c$ are
lattice constants the same as those in BC-Tab. 3.1. \label{tab:umax}}

{\renewcommand{\arraystretch}{1.4}
\begin{centering}
\begin{tabular}{ccccccc}
\hline 
Type of BZ & \multicolumn{6}{c}{$u_{{\rm max}}$ for HS lines}\tabularnewline
\hline 
\lstinline!"OrthBase(a)"! & $\Delta\text{: }\frac{1}{4}+\frac{b^{2}}{4a^{2}}$ & $F\text{: }\frac{1}{4}-\frac{b^{2}}{4a^{2}}$ & $B\text{: }\frac{1}{4}+\frac{b^{2}}{4a^{2}}$ & $G\text{: }\frac{1}{4}-\frac{b^{2}}{4a^{2}}$ &  & \tabularnewline
\lstinline!"OrthBase(b)"! & $A\text{: }\frac{1}{4}+\frac{a^{2}}{4b^{2}}$ & $E\text{: }\frac{1}{4}-\frac{a^{2}}{4b^{2}}$ & $\Sigma\text{: }\frac{1}{4}+\frac{a^{2}}{4b^{2}}$ & $C\text{: }\frac{1}{4}-\frac{a^{2}}{4b^{2}}$ &  & \tabularnewline
\lstinline!"OrthBody(a)"! & $\Lambda\text{: }\frac{1}{4}+\frac{c^{2}}{4a^{2}}$ & $G\text{: }\frac{1}{4}-\frac{c^{2}}{4a^{2}}$ & $\Delta\text{: }\frac{1}{4}+\frac{b^{2}}{4a^{2}}$ & $U\text{: }\frac{1}{4}-\frac{b^{2}}{4a^{2}}$ &  & \tabularnewline
\lstinline!"OrthBody(b)"! & $\Lambda\text{: }\frac{1}{4}+\frac{c^{2}}{4b^{2}}$ & $G\text{: }\frac{1}{4}-\frac{c^{2}}{4b^{2}}$ & $\Sigma\text{: }\frac{1}{4}+\frac{a^{2}}{4b^{2}}$ & $F\text{: }\frac{1}{4}-\frac{a^{2}}{4b^{2}}$ &  & \tabularnewline
\lstinline!"OrthBody(c)"! & $\Delta\text{: }\frac{1}{4}+\frac{b^{2}}{4c^{2}}$ & $U\text{: }\frac{1}{4}-\frac{b^{2}}{4c^{2}}$ & $\Sigma\text{: }\frac{1}{4}+\frac{a^{2}}{4c^{2}}$ & $F\text{: }\frac{1}{4}-\frac{a^{2}}{4c^{2}}$ &  & \tabularnewline
\lstinline!"OrthFace(a)"! & $G\text{: }\frac{1}{4}+C_{ab}^{-}$ & $H\text{: }\frac{1}{4}-C_{ab}^{-}$ & $C\text{: }\frac{1}{4}+A_{bc}^{-}$ & $A\text{: }\frac{1}{4}-A_{bc}^{-}$ & $D\text{: }\frac{1}{4}+B_{ac}^{-}$ & $B\text{: }\frac{1}{4}-B_{ac}^{-}$\tabularnewline
\lstinline!"OrthFace(b)"! & $\text{\ensuremath{\Lambda}}\text{: }\frac{1}{4}+C_{ab}^{+}$ & $Q\text{: }\frac{1}{4}-C_{ab}^{+}$ & $G\text{: }\frac{1}{4}+C_{ab}^{-}$ & $H\text{: }\frac{1}{4}-C_{ab}^{-}$ & \multicolumn{2}{c}{$(C_{ab}^{\pm}=\frac{c^{2}(a^{2}\pm b^{2})}{4a^{2}b^{2}})$}\tabularnewline
\lstinline!"OrthFace(c)"! & $\Delta\text{: }\frac{1}{4}+B_{ac}^{+}$ & $R\text{: }\frac{1}{4}-B_{ac}^{+}$ & $D\text{: }\frac{1}{4}+B_{ac}^{-}$ & $B\text{: }\frac{1}{4}-B_{ac}^{-}$ & \multicolumn{2}{c}{$(B_{ac}^{\pm}=\frac{b^{2}(a^{2}\pm c^{2})}{4a^{2}c^{2}})$}\tabularnewline
\lstinline!"OrthFace(d)"! & $\Sigma\text{: }\frac{1}{4}+A_{bc}^{+}$ & $U\text{: }\frac{1}{4}-A_{bc}^{+}$ & $C\text{: }\frac{1}{4}+A_{bc}^{-}$ & $A\text{: }\frac{1}{4}-A_{bc}^{-}$ & \multicolumn{2}{c}{$(A_{bc}^{\pm}=\frac{a^{2}(b^{2}\pm c^{2})}{4b^{2}c^{2}})$}\tabularnewline
\lstinline!"TetrBody(a)"! & $\Lambda\text{: }\frac{1}{4}+\frac{c^{2}}{4a^{2}}$ & $V\text{: }\frac{1}{4}-\frac{c^{2}}{4a^{2}}$ &  &  &  & \tabularnewline
\lstinline!"TetrBody(b)"! & $\Sigma\text{: }\frac{1}{4}+\frac{a^{2}}{4c^{2}}$ & $F\text{: }\frac{1}{4}-\frac{a^{2}}{4c^{2}}$ & $U\text{: }\frac{1}{2}-\frac{a^{2}}{2c^{2}}$ & $Y\text{: }\frac{a^{2}}{2c^{2}}$ &  & \tabularnewline
\lstinline!"TrigPrim(a)"! & $\Lambda\text{: }\frac{1}{6}+\frac{2c^{2}}{3a^{2}}$ & $P\text{: }\frac{1}{3}-\frac{2c^{2}}{3a^{2}}$ &  &  &  & \tabularnewline
\lstinline!"TrigPrim(b)"! & $B\text{: }\frac{1}{3}-\frac{a^{2}}{6c^{2}}$ & $Y\text{: }\frac{1}{6}+\frac{a^{2}}{6c^{2}}$ & $\Sigma\text{: }\frac{1}{3}+\frac{a^{2}}{12c^{2}}$ & $Q\text{: }\frac{1}{6}-\frac{a^{2}}{12c^{2}}$ &  & \tabularnewline
\lstinline!"HexaPrim"! & $T\text{: }\frac{1}{3}$ & $T'\text{: }\frac{1}{6}$ & $S\text{: }\frac{1}{3}$ & $S'\text{: }\frac{1}{6}$ &  & \tabularnewline
\lstinline!"CubiBody"! & $\Lambda\text{: }\frac{1}{4}$ & $F:\frac{1}{4}$ &  &  &  & \tabularnewline
\lstinline!"CubiFace"! & $\Sigma\text{: }\frac{3}{8}$ & $S:\frac{1}{8}$ &  &  &  & \tabularnewline
\hline 
\end{tabular}
\par\end{centering}
}
\end{table}

The information about a k-point from \lstinline!identifyBCHSKpt!
is preliminary. In the final result, the $\{S|\bm{w}\}$ defined in
subsection \ref{subsec:LGIR-at-any-k} is determined and only one
kname is determined with the knowledge of actual values of lattice
constants or basic vectors. This is done by \lstinline!identifyBCHSKptBySG!
which returns the complete information of a k-point in the list form
of 
\begin{equation}
\{\vk,\text{\,kname},\,\text{\text{line\_info},\,}\bar{G}^{\vk},\,\vk_{{\rm BC}},\,\{S|\bm{w}\},\,S\vk_{{\rm BC}},\,\bm{g},\ {\rm u}\rightarrow v_{{\rm u}},\,u_{{\rm max}},\,\text{if\/inG}\}\label{eq:kinfo}
\end{equation}
in which $\bar{G}^{\vk}$ is the little co-group, $u_{{\rm max}}$
is the maximum value of $u$ that keeps $\vk_{{\rm BC}}$ inside or
on the boundary of the BZ, and if\/inG is a string ``in G'' or
``not in G'' to indicate whether $\{S|\bm{w}\}\in G$ or not. In
the above example, in order to determine whether the k-point $(-0.3,-0.3,0.3)$
is $\Lambda$ or $V$, the $u_{{\rm max}}$ of $\Lambda$ and $V$
should be known and then the one satisfying $v_{{\rm u}}<u_{{\rm max}}$
is selected. We can see that $u_{{\rm max}}$ depends on the actual
values of lattice constants in most cases listed in Tab. \ref{tab:umax}.
Therefore, the basic vectors are needed to precisely determine $\Lambda$
or $V$. Taking the space group $I4_{1}md$ (No. 109) for example,
it has body-centered tetragonal lattice and has (a) type BZ when $a>c$.
If $a=3$, $c=2$ then $u_{\max}^{\Lambda}=0.361111$ and $u_{\max}^{V}=0.138889$,
which makes $0.3<u_{{\rm max}}^{\Lambda}$, $0.2>u_{{\rm max}}^{V}$
and then $\Lambda$ is selected. If $a=3$, $c=1$ then $u_{\max}^{\Lambda}=0.222222$
and $u_{\max}^{V}=0.277778$, which makes $0.3>u_{{\rm max}}^{\Lambda}$,
$0.2<u_{{\rm max}}^{V}$ and then $V$ is selected. The code is as
follow.

\begin{lstlisting}[backgroundcolor={\color{yellow!5!white}},mathescape=true]
|In[1]:=| bv=BasicVectors["TetrBody"]
        identifyBCHSKptBySG[109,bv/.{a->3,c->2},{-0.3,-0.3,0.3}]
        identifyBCHSKptBySG[109,bv/.{a->3,c->1},{-0.3,-0.3,0.3}]
|Out[1]=| {{-a/2,a/2,c/2},{a/2,-a/2,c/2},{a/2,a/2,-c/2}}
|Out[2]=| {{-0.3,-0.3,0.3},$\Lambda$,$\Gamma\Lambda$,C4v,{u,u,-u},{C2x,{3/4,1/4,1/2}},
         {-u,-u,u},{0,0,0},u->0.3,0.361111,not in G}
|Out[3]=| {{-0.3,-0.3,0.3},V,ZV,C4v,{-1/2+u,1/2+u,1/2-u},{E,{0,0,0}},
         {-1/2+u,1/2+u,1/2-u},{0,-1,0},u->0.2,0.222222,in G}
\end{lstlisting}
When the basic vectors are unknown, we can use the \lstinline!BZtype!,
i.e. \lstinline!"a"!, \lstinline!"b"!, \lstinline!"c"!, \lstinline!"d"!,
\lstinline!""!, as the second parameter of \lstinline!identifyBCHSKptBySG!.
In this case, k-points on the HS lines listed in Tab. \ref{tab:umax}
except the last three rows may be identified with incorrect names,
because the $u_{{\rm max}}$ cannot be determined precisely without
actual values of $a,b,c$ and in this case $u_{{\rm max}}$ is set
to a makeshift value $1/4$ to make sure only one kname is output.
However, if $u_{{\rm max}}$ is not precisely determined and the option
\lstinline!"allowtwok"->True! is used, \lstinline!identifyBCHSKptBySG!
can still return two entries of k-point information with different
knames for k-points on the HS lines in Tab. \ref{tab:umax}.

\subsection{Tables for LG IRs}

\begin{figure}
\begin{centering}
\includegraphics[width=12cm]{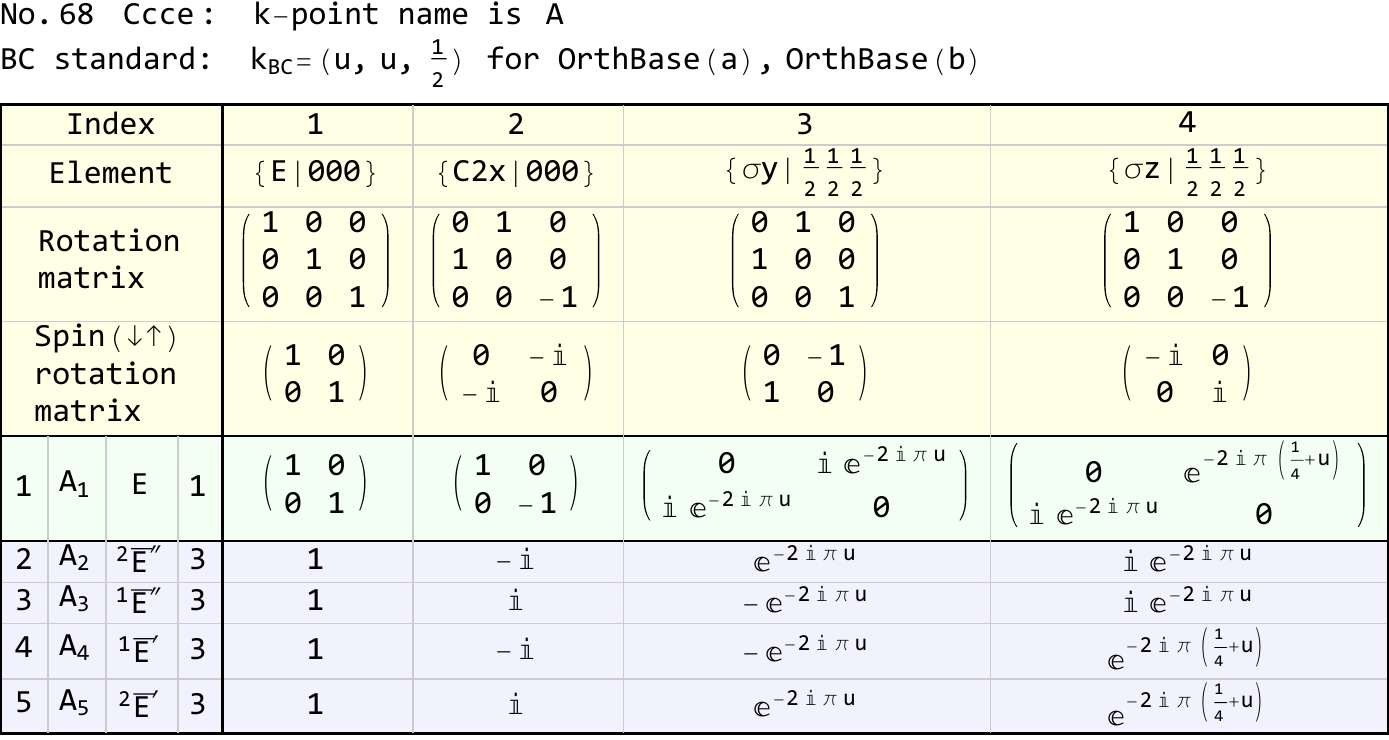}
\par\end{centering}
\caption{The output of the function \lstinline!showLGIrepTab[68,"A"]! which
shows the table for LG IRs of the k-point $A$ of the space group
of number 68. Light green background shows single-valued IRs and light
blue background shows double-valued IRs. Both the $\Gamma$ label
(the second column) and the extended Mulliken label (the third column)
are given. The first column is the index of the LG IRs and the fourth
column is the realities whose values may be 1, 2, 3, or x. \label{fig:IR68A}}
\end{figure}
\begin{figure}[t]
\begin{centering}
\includegraphics[width=12cm]{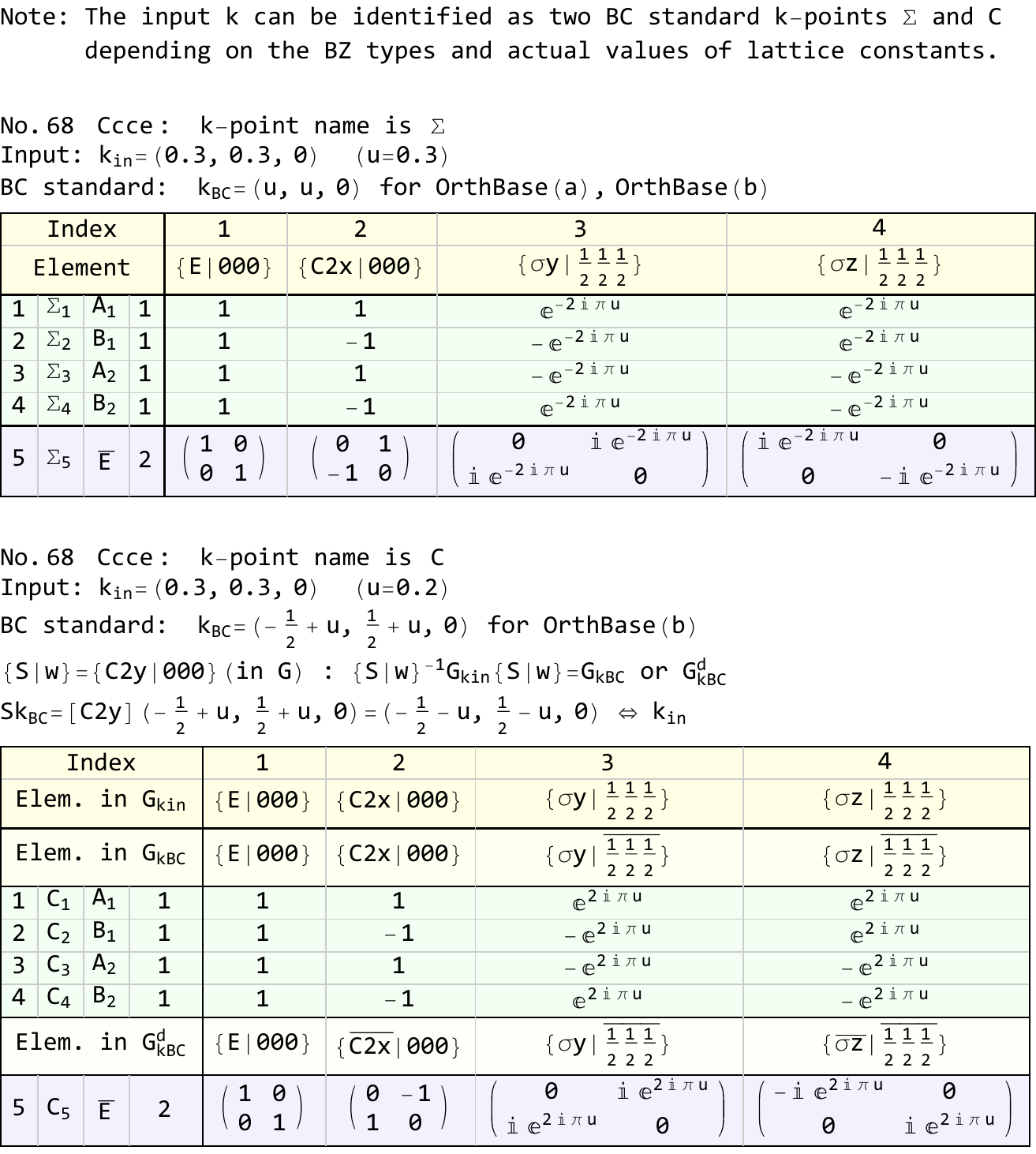}
\par\end{centering}
\caption{The output of the function \lstinline!showLGIrepTab[68,\{0.3,0.3,0\},"rotmat"->False]!
which shows two tables for LG IRs of the k-point $(0.3,0.3,0)$ of
the space group of number 68. In this example $(0.3,0.3,0)$ is identified
as two k-points $\Sigma$ and $C$, and which one is selected depends
on the actual values of lattice constants. $k_{{\rm in}}$ means the
input k-point, $G_{{\rm kin}}$ means the little group of $k_{{\rm in}}$,
$G_{{\rm kBC}}$ means the little group of the standard BC k-point
$k_{{\rm BC}}$, and $G_{{\rm kBC}}^{{\rm d}}$ means the double little
group of $k_{{\rm BC}}.$ \label{fig:IR68-SM-C}}
\end{figure}

Although the data of LG IRs are contained in the BC book, they are
distributed over several tables. This makes it not direct to obtain
the representation matrices or characters of a designated LG IR, which
is shown by the aforementioned example in the introduction. Accordingly,
we create the functions \lstinline!getLGIrepTab[sgno,k]! and \lstinline!showLGIrepTab[sgno,k]!
in which \lstinline!k! is either the name or the numeric coordinates
of a k-point. The former calculates and gives the representation matrices
and related information of all the single-valued and double-valued
LG IRs of \lstinline!k! for the space group of number \lstinline!sgno!,
and the latter shows the data in user-friendly table form. 

Fig. \ref{fig:IR68A} is an example showing the output of \lstinline!showLGIrepTab[68,"A"]!
which directly gives the information about the k-point, the available
types of BZ, the LG elements (only the coset representatives with
respect to the translation group are given), the rotation matrices
and spin rotation matrices, the representation matrices of both single-valued
LG IRs (light green background) and double-valued LG IRs (light blue
background), the $\Gamma$ labels ($A_{1},A_{2},\cdots$) and extended
Mulliken labels ($E,$ $^{2}\!\bar{E}'',\cdots$) of LG IRs, and the
realities (the fourth column) of the corresponding SG IRs. Following
the notations in the BC book, the realities 1, 2, 3 stand for real
representation, pseudo-real representation, and complex representation
respectively for the SG IRs in which $\vk$ and $-\vk$ are in the
same star. If $\vk$ and $-\vk$ are not in the same star, the SG
IR is complex and its reality is represented by a letter ``x''.
For double-valued LG IRs, the representation matrix of the element
$\{\bar{R}|\bm{v}\}$ is just the negative value of the representation
matrix of $\{R|\bm{v}\},$ so only $\{R|\bm{v}\}$'s are shown in
the LG IR table in Fig. \ref{fig:IR68A}. 

If the coordinates of a k-point are given, they may be identified
as two BC standard k-points for the k-points listed in Tab. \ref{tab:umax},
and two LG IR tables are given by \lstinline!showLGIrepTab!. An example
of this case is \lstinline!showLGIrepTab[68,{0.3,0.3,0},"rotmat"->False]!
whose output is shown in Fig. \ref{fig:IR68-SM-C}. The input k-point
$\vk_{{\rm in}}=(0.3,0.3,0)$ is identified as only $\Sigma$ for
\lstinline!"OrthBase(a)"! type of BZ but as either $\Sigma$ or $C$
for \lstinline!"OrthBase(b)"! type of BZ. Two LG IR tables for both
$\Sigma$ and $C$ are given in Fig. \ref{fig:IR68-SM-C}. In this
example, the coordinates $(0.3,0.3,0)$ are directly in the form of
$(u,u,0)$, i.e. the coordinates of the BC standard $\Sigma$ k-point,
and hence $\vk_{{\rm in}}$ is of type I if it is a $\Sigma$ k-point.
But if this k-point is identified as $C$, it is of type II and has
non-identity $\{S|\bm{w}\}$ which can relate this k-point to the
BC standard $C$ k-point. Here $\{S|\bm{w}\}=\{C_{2y}|000\}$ and
it makes the little group of $\vk_{{\rm in}}$, $G_{{\rm kin}}$,
isomorphic to the (double) little group of $\vk_{{\rm BC}}$, $G_{{\rm kBC}}$
($G_{{\rm kBC}}^{{\rm d}}$), i.e. 
\begin{equation}
\{S|\bm{w}\}^{-1}G_{{\rm kin}}\{S|\bm{w}\}=G_{{\rm kBC}}\ \ \text{or }G_{{\rm kBC}}^{{\rm d}}.\label{eq:GkinGkBC}
\end{equation}
Then the LG IRs of $\vk_{{\rm in}}$ are obtained from those of $\vk_{{\rm BC}}$
according to Eqs. (\ref{eq:GMkGMkBC}) and (\ref{eq:GkinGkBC}). The
mappings between the elements in $G_{{\rm kin}}$ and those in $G_{{\rm kBC}}$
($G_{{\rm kBC}}^{{\rm d}}$) are also shown in the lower table of
Fig. \ref{fig:IR68-SM-C}, which can help to understand the relations
between the LG IRs of $\vk_{{\rm in}}$ and $\vk_{{\rm BC}}.$ Let's
remind that here $\vk_{{\rm in}}$ just borrows the name $C$ of $\vk_{{\rm BC}}$
and if we give a different name to $\vk_{{\rm in}},$ say $C'$, then
the LG IR labels of $\vk_{{\rm in}}$ should be $C_{1}',\cdots,C_{5}'$
which are clearly distinguished from $C_{1},\cdots,C_{5}$ of $\vk_{{\rm BC}}.$

Fig. \ref{fig:IR68-SM-C} is generated by \lstinline!showLGIrepTab!
with the option \lstinline!"rotmat"->False! which controls not to
show the rotation matrices. In fact, \lstinline!showLGIrepTab! has
several options. Its default options can be obtained by 
\begin{lstlisting}[backgroundcolor={\color{yellow!5!white}},mathescape=true]
|In[1]:=| Options[showLGIrepTab]//InputForm
|Out[1]=| {"uNumeric"->False, "irep"->All, "elem"->All, "rotmat"->True, 
         "trace"->False, "spin"->"downup", "abcOrBasVec"->None, 
         "linewidth"-> 2}
\end{lstlisting}
If \lstinline!"uNumeric"->True! is used, the value of $u$ is substituted
into $u$ to make the LG IRs numeric. Although the LG IRs of $\Sigma$
and $C$ in Fig. \ref{fig:IR68-SM-C} are different seemingly, it
will be seen clearly that they are in fact equivalent when \lstinline!"uNumeric"->True!
is used, with the correspondence $\Sigma_{1,3}\leftrightarrow C_{3,1}$,
$\Sigma_{2,4}\leftrightarrow C_{4,2}$, $\Sigma_{5}\leftrightarrow C_{5}.$
Options \lstinline!"irep"! and \lstinline!"elem"! can select certain
IRs and elements to be shown, e.g. \lstinline!"irep"->{1,3},"elem"->{3,4}!
will only show the first and third LG IRs ($\Sigma_{1,3}$ and $C_{1,3}$
in Fig. \ref{fig:IR68-SM-C}) and the third and fourth elements ($\{\sigma_{y}|\frac{1}{2}\frac{1}{2}\frac{1}{2}\}$
and $\{\sigma_{z}|\frac{1}{2}\frac{1}{2}\frac{1}{2}\}$ in Fig. \ref{fig:IR68-SM-C}).
\lstinline!"trace"->True! makes \lstinline!showLGIrepTab! show the
characters not the representation matrices. In fact, we have also
defined functions \lstinline!getLGCharTab[sgno,k]! and \lstinline!showLGCharTab[sgno,k]!
to calculate and show the character tables for LG IRs and these two
functions are just respectively the functions \lstinline!getLGIrepTab[sgno,k]!
and \lstinline!showLGIrepTab[sgno,k]! with the option \lstinline!"trace"->True!.
The option \lstinline!"spin"->"updown"! will change the bases of
spin rotation matrices from the default $\{\downarrow,\uparrow\}$
to $\{\uparrow,\downarrow\}$. If lattice constants or basic vectors
are given through the option \lstinline!"abcOrBasVec"!, one definite
k-point is determined, e.g. \lstinline!"abcOrBasVec"->{a->3,b->5,c->4}!
only shows $\Sigma$ and \lstinline!"abcOrBasVec"->{a->2,b->5,c->4}!
only shows $C$ for the example in Fig. \ref{fig:IR68-SM-C}. At last,
\lstinline!"linewidth"! can control the line width of the table.

\subsection{Tables for SG IRs}

\begin{figure}
\begin{centering}
\includegraphics[width=15cm]{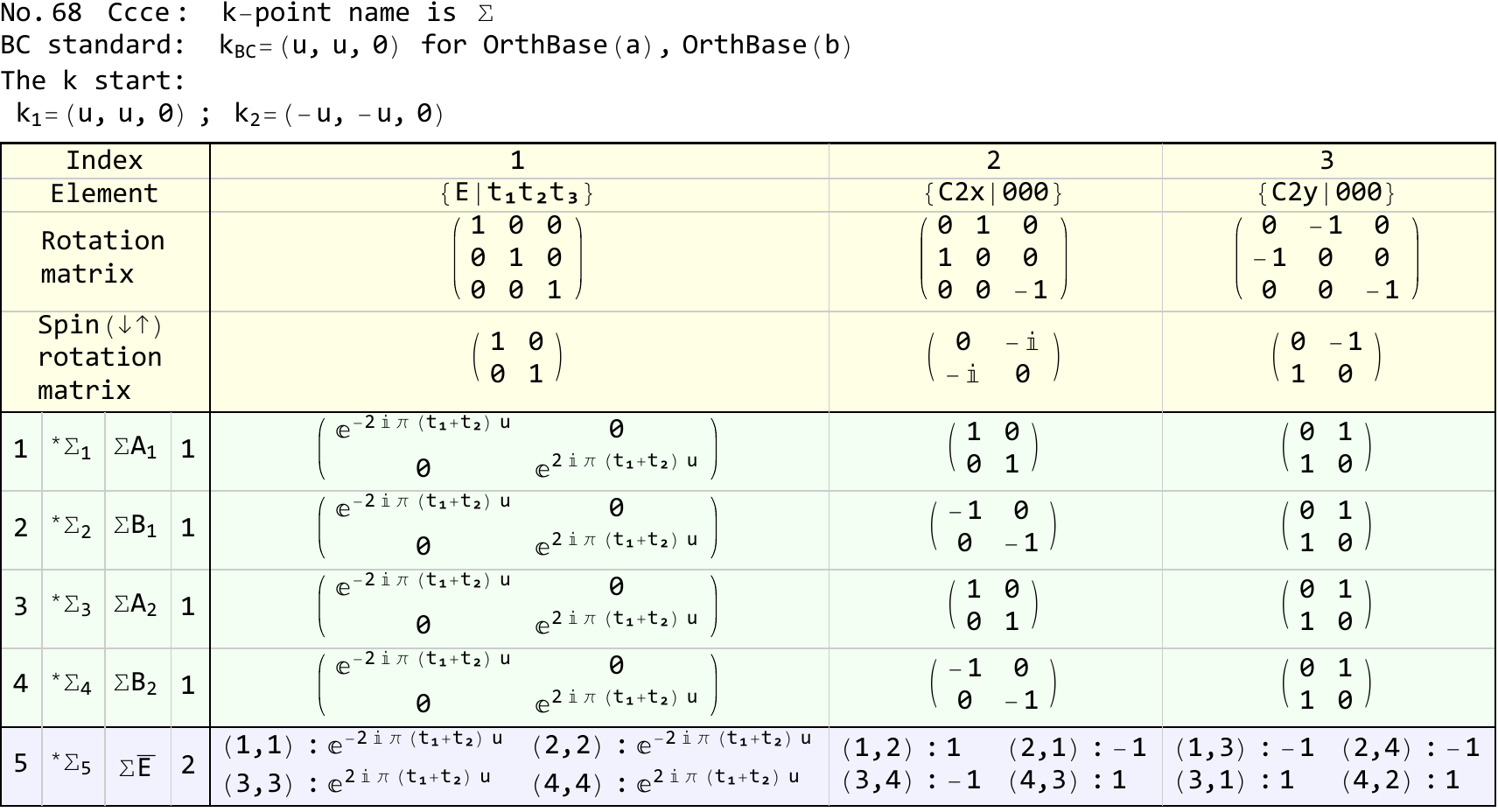}
\par\end{centering}
\caption{The output of \lstinline[mathescape=true]!showSGIrepTab[68,"$\Sigma$","maxDim"->3,"elem"->\{1,2,3\}]!
which shows part of the table for SG IRs of the star $^{*}\Sigma$
of the space group of number 68. Light green background shows single-valued
IRs and light blue background shows double-valued IRs. The first column
is the index of the SG IRs, the second and third column are two kinds
of labels for SG IRs, and the fourth column is the realities whose
values may be 1, 2, 3, or x. \label{fig:SGIR68SM}}
\end{figure}

In addition to tables for LG IRs, we have also created the functions
\lstinline!getSGIrepTab[sgno,k]! and \lstinline!showSGIrepTab[sgno,k]!
to calculate and show the tables for SG IRs. The usage of \lstinline!showSGIrepTab[sgno,k]!
is almost the same as \lstinline!showLGIrepTab[sgno,k]! except that
\lstinline!showSGIrepTab! has one more option \lstinline!"maxDim"!.
\lstinline!"maxDim"! is the critical dimension controlling the appearance
of representations whose default value is 4. When the dimension of
a representation matrix is lower than or equal to \lstinline!"maxDim"!
, the representation matrix is shown in matrix form, and otherwise
only nonzero matrix elements are shown to save the table space. An
example is shown in Fig. \ref{fig:SGIR68SM}, which is generated by
the following code.
\begin{lstlisting}[backgroundcolor={\color{yellow!5!white}},mathescape=true]
showSGIrepTab[68, "$\Sigma$", "maxDim"->3, "elem"->{1,2,3}]
\end{lstlisting}
This example gives the table for the SG IRs of the wave vector star
$^{*}\Sigma$ of space group 68. The double-valued SG IR $^{*}\Sigma_{5}$
is shown in the form of nonzero matrix elements, because its dimension
4 is larger than the value 3 of \lstinline!"maxDim"!. To save space,
representation matrices of only the first three SG elements are shown
due to the option \lstinline!"elem"->{1,2,3}!. If the option \lstinline!"elem"!
is not used, the table will show the representation matrices of 8
SG elements in total, i.e. all the coset representatives with respective
to the translation group. We use two kinds of labels for SG IRs. The
first one is to put a $*$ at the top left corner of corresponding
$\Gamma$ label of LG IR, and the second one is to put the k-point
name in front of the extended Mulliken label of LG IR, e.g. both $^{*}\Sigma_{2}$
and $\Sigma B_{1}$ for $\Sigma_{2}\uparrow G$. 

\section{Direct product of SG IRs}

The (inner) direct product of SG IRs is of great importance to determine
the selection rules in various quantum processes in crystals. Therefore
we have realized the decomposition (or reduction) of the direct product
of any two SG IRs according to BC-Eqs. (4.7.1) and (4.7.29), i.e.
\begin{equation}
(\Gamma_{p}^{i}\uparrow G)\otimes(\Gamma_{q}^{j}\uparrow G)\equiv\sum_{l}\sum_{r}C_{pq,r}^{ij,l}(\Gamma_{r}^{l}\uparrow G)
\end{equation}
\begin{equation}
C_{pq,r}^{ij,l}=\sum_{\{\alpha|\bm{u}\}}\!\!\vphantom{\sum}'\sum_{\{\beta|\bm{v}\}}\!\!\vphantom{\sum}'\frac{|T|}{|N_{\alpha\beta}|}\sum_{\{\gamma|\bm{w}\}\in N_{\alpha\beta}/T}\chi_{p}^{i}(\{\beta|\bm{v}\}^{-1}\{\gamma|\bm{w}\}\{\beta|\bm{v}\})\cdot\chi_{q}^{j}(\{\alpha|\bm{u}\}\{\gamma|\bm{w}\}\{\alpha|\bm{u}\})\cdot\chi_{r}^{l*}(\{\gamma|\bm{w}\})
\end{equation}
in which $\sum\vphantom{\sum}'$ means the summation is restricted
by the condition 
\begin{equation}
\beta\vk_{i}+\alpha\vk_{j}\equiv\vk_{l}.
\end{equation}
In the above equations, $\Gamma_{p}^{i}$, $\Gamma_{q}^{j},$ and
$\Gamma_{r}^{l}$ are the LG IRs of the little groups $G^{\vk_{i}}$,
$G^{\vk_{j}},$ and $G^{\vk_{l}}$ respectively; $\chi_{p}^{i},$
$\chi_{q}^{j},$ and $\chi_{r}^{l}$ are their characters respectively;
and $\Gamma_{p}^{i}\uparrow G$, $\Gamma_{q}^{j}\uparrow G$, and
$\Gamma_{r}^{l}\uparrow G$ are the corresponding induced SG IRs of
the space group $G$. $\{\alpha|\bm{u}\}$'s are the double coset
representatives of $G$ with respect to $G^{\vk_{l}}$ and $G^{\vk_{j}}$;
$\{\beta|\bm{v}\}$'s are the double coset representatives of $G$
with respect to $L_{\alpha}=G^{\vk_{l}}\cap(\{\alpha|\bm{u}\}G^{\vk_{j}}\{\alpha|\bm{u}\}^{-1})$
and $G^{\vk_{i}}$; and $N_{\alpha\beta}$ is a group defined by $N_{\alpha\beta}=L_{\alpha}\cap(\{\beta|\bm{v}\}G^{\vk_{i}}\{\beta|\bm{v}\}^{-1})$. 

\begin{figure}[t]
\begin{centering}
\includegraphics[width=12cm]{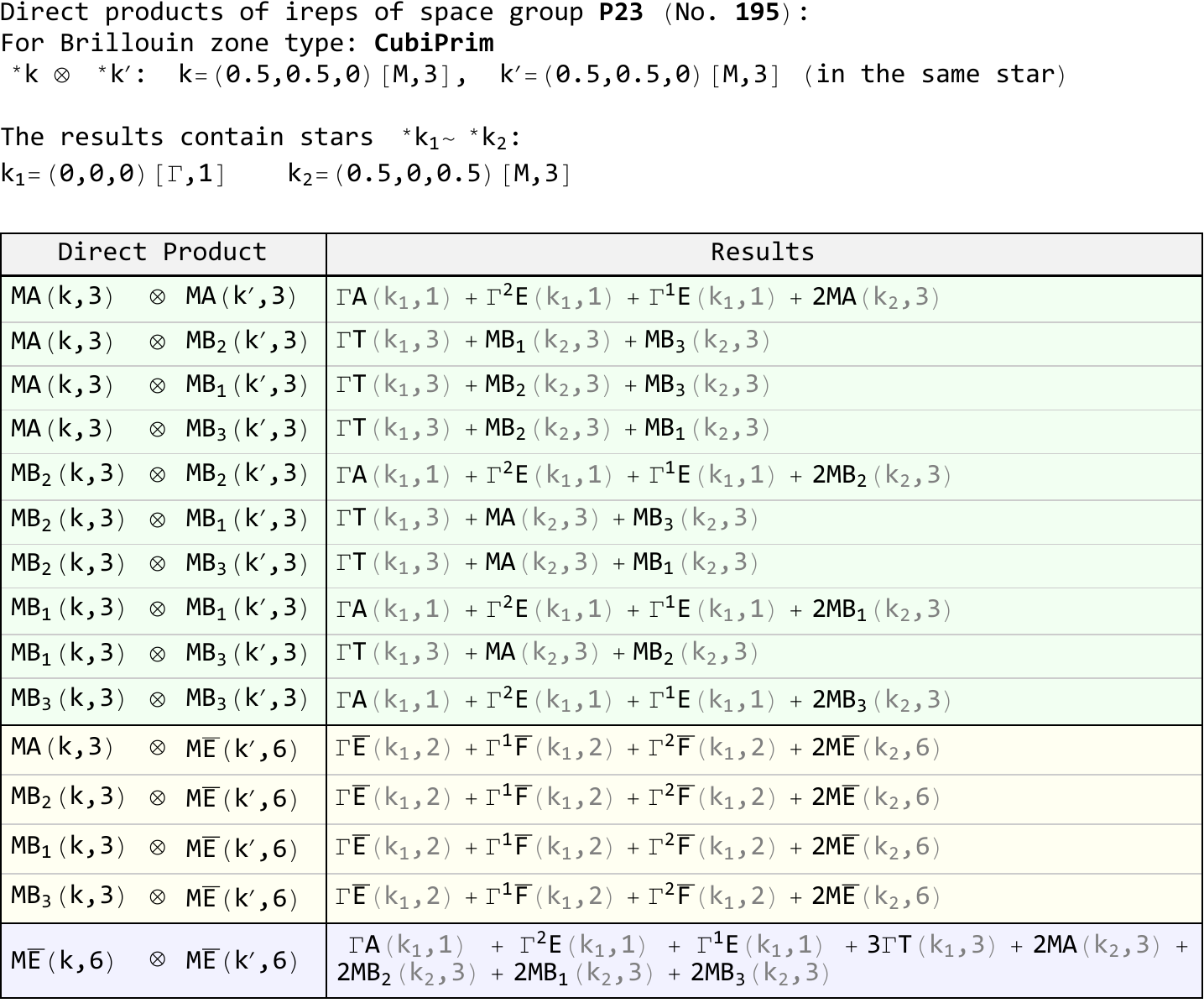}
\par\end{centering}
\caption{The output of \lstinline!showSGIrepDirectProduct[195, \{1/2,1/2,0\}, \{1/2,1/2,0\}, "label"->2]!
which shows the direct product of SG IRs of $^{*}M$ and $^{*}M$
for space group $P23$. The notation $[M,3]$ on top of the table
means the the k-point in front of it is identified as $M$ and the
number of arms of its star is 3. The notation $(k_{2},3)$ in the
table indicates that the SG IR in front of it is for the star of $k_{2}$
and the dimension of this SG IR is 3. Light green background stands
for direct products between single-valued SG IRs and single-valued
SG IRs, light yellow background for direct products between single-valued
SG IRs and double-valued SG IRs, and light blue background for direct
products between double-valued SG IRs and double-valued SG IRs.\label{fig:SGIRDP-MM}}
\end{figure}

The functions that calculate and show the direct product of SG IRs
are \lstinline!SGIrepDirectProduct[sgno,k1,k2]! and \lstinline!showSGIrepDirectProduct[sgno,k1,k2]!
respectively, in which \lstinline!k1! and \lstinline!k2! are both
numeric k-point coordinates. Taking the same example as in the BC
book, that is the direct products of SG IRs of $^{*}M$ and $^{*}M$
for space group $P23$ (No. 195). The following function 
\begin{lstlisting}[backgroundcolor={\color{yellow!5!white}},mathescape=true]
showSGIrepDirectProduct[195, {1/2,1/2,0}, {1/2,1/2,0}, "label"->2]
\end{lstlisting}
gives the results shown in Fig. \ref{fig:SGIRDP-MM}, which are consistent
with the direct products listed in BC-Tab. 4.5. \lstinline!showSGIrepDirectProduct!
has three options, \lstinline!"label"!, \lstinline!"abcOrBasVec"!,
and \lstinline!"linewidth"!, in which \lstinline!"label"! whose
value can be 1 (default) or 2 controls which kind of labels for SG
IRs are used and the other two options have the same usages as in
\lstinline!showLGIrepTab!. It is worth noting that \lstinline!showSGIrepDirectProduct!
not only gives the direct products between single-valued SG IRs and
single-valued SG IRs, but also gives the direct products between double-valued
SG IRs and double-valued SG IRs and even direct products between single-valued
SG IRs and double-valued SG IRs.

\section{Obtain the LG IRs of energy bands}

As mentioned in the introduction, a tool with full support for determining
the LG IRs of all Bloch states in energy bands has been missing for
a long time until the appearance of the recent program \textsf{irvsp}\citep{Gao_Wang_2020___2002.04032v1_Irvsp}.
However, \textsf{irvsp} only supports LG IRs in the BCS convention.
Therefore, support for LG IRs in the BC convention is provided in
the \textsf{SpaceGroupIrep} package, as a complement to \textsf{irvsp}.

Here, we use ``BC cells'' for the primitive cells with BC settings,
i.e. the cells have basic vectors defined in BC-Tab. 3.1 and the SG
symmetry operations defined by the cells are consistent with the first-row
generators of every space group in BC-Tab. 3.7. Note that a cell only
with basic vectors defined in BC-Tab. 3.1 is not necessary a BC cell,
because different selections for the origin may result in different
SG elements. In the following two subsections, we first discuss the
cases with BC cells, and then discuss the cases with non-BC cells.

\subsection{For BC cells}

To determine the LG IRs of energy bands, the character of each LG
element operating on each set of degenerate Bloch states should be
obtained first. Fortunately, there has be such a program called \textsf{vasp2trace}\citep{Vergniory_Wang_2019_566_480__complete}
which can do this. \textsf{vasp2trace} is a third-party post-processing
program for \textsf{VASP}. It reads the output wave functions of \textsf{VASP},
calculates the characters of LG elements, and writes the results in
a file named \textsf{trace.txt}. In fact, \textsf{vasp2trace} is the
precursor of \textsf{irvsp} without the function of determining LG
IRs and \textsf{irvsp} can also output a \textsf{trace.txt} file in
certain cases. However, the\textsf{ trace.txt} files generated by
the two programs may be different and what we need is the \textsf{trace.txt
}generated by \textsf{vasp2trace}, because the chacracter data in
the \textsf{trace.txt }file from\textsf{ irvsp} may be processed data,
not the original ones we need. It is worth noting that the number
of bands output by \textsf{vasp2trace} is limited by the number of
electrons, i.e. the NELECT in \textsf{VASP}, because \textsf{vasp2trace}
is designed for determining the topological properties of materials
and for this purpose only the trace data of occupied states are needed.
To make \textsf{vasp2trace} output trace data for all bands, two tiny
changes should be made to the source code of \textsf{vasp2trace}:
changing the \lstinline!nele! in the 30th line of \textsf{wrtir.f90}
to \lstinline!ne! and deleting the 55th line of \textsf{chrct.f90},
i.e. \lstinline!IF(IE>nele) EXIT!.

\begin{figure}[ph]
\begin{centering}
\includegraphics[width=12cm]{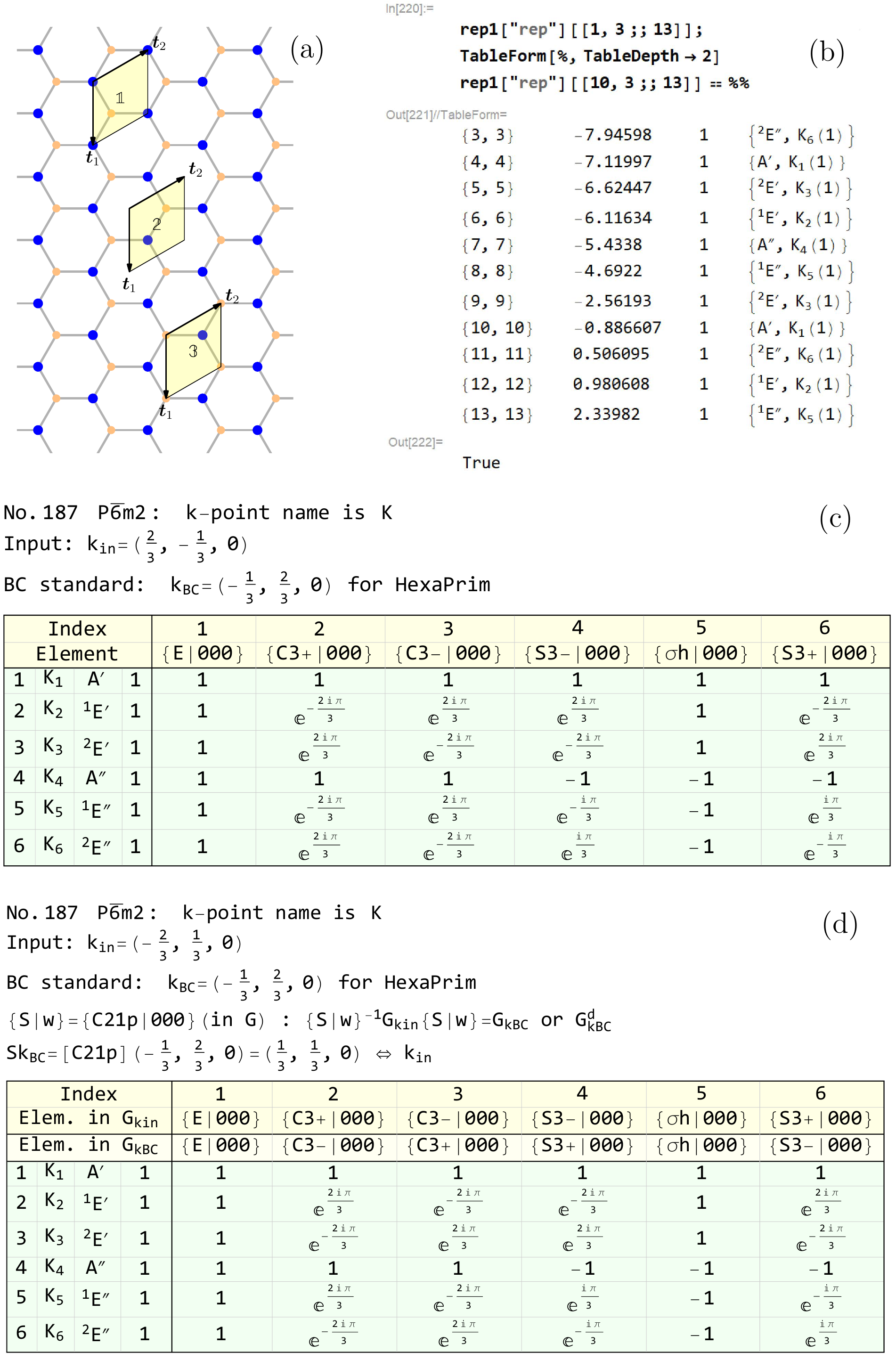}
\par\end{centering}
\caption{The LG IRs of monolayer MoS$_{2}$. (a) Top view and three different
BC cells for monolayer MoS$_{2}$ where blue and orange dots represent
Mo and S atoms respectively. (b) Codes to show the LG IRs at $K$
(the first k-point) and $K'$ (the 10th k-point) for seven valence
bands (3\textendash 9) and four conduction bands (10\textendash 13).
The four columns of the output are respectively the range of bands
for a degenerate energy (e.g. \{3,3\} means from band 3 to band 3),
the band energy, the degree of degeneracy, and the labels for LG IRs.
Both the extended Mulliken label and the $\Gamma$ label for a LG
IR are given and the integer in the parentheses is the dimension of
the LG IR. (c) The character table for LG IRs of $K$ $(\frac{2}{3},-\frac{1}{3},0)$
obtained by \lstinline!showLGCharTab[187,\{2/3,-1/3,0\},"rotmat"->False,"irep"->1;;6]!.
(d) The character table for LG IRs of $K'$ $(-\frac{2}{3},\frac{1}{3},0)$
obtained by \lstinline!showLGCharTab[187,\{-2/3,1/3,0\},"rotmat"->False,"irep"->1;;6]!.
\label{fig:MoS2}}
\end{figure}

In the \textsf{SpaceGroupIrep} package we use the function \lstinline!readVasp2trace!
to read the \textsf{trace.txt }file generated by \textsf{vasp2trace}
and its returned value is an association containing all the data in
the \textsf{trace.txt }file. Then the function \lstinline!getBandRep!
is used to determine the LG IRs according to the trace data returned
by \lstinline!readVasp2trace!. There are three ways to call \lstinline!getBandRep!,
i.e.
\begin{lstlisting}[backgroundcolor={\color{yellow!5!white}},mathescape=true]
getBandRep[sgno, BZtypeOrBasVec, traceData]
getBandRep[sgno, BZtypeOrBasVec, traceData, ikOrListOrSpan]
getBandRep[sgno, BZtypeOrBasVec, traceData, ikOrListOrSpan, ibOrListOrSpan]
\end{lstlisting}
in which \lstinline!BZtypeOrBasVec! has the same meaning as in \lstinline!identifyBCHSKptBySG!,
\lstinline!traceData! is the value returned by \lstinline!readVasp2trace!,
\lstinline!ikOrListOrSpan! (\lstinline!ibOrListOrSpan!) specifies
the indexes of k-points (bands) to be processed. \lstinline!ikOrListOrSpan!
(\lstinline!ibOrListOrSpan!) may be an integer such as \lstinline!5!,
a list of integers such as \lstinline!{2,3,5}!, or a span such as
\lstinline!3;;5!. If \lstinline!ikOrListOrSpan! (\lstinline!ibOrListOrSpan!)
is not specified then all k-points (bands) are processed. Taking monolayer
MoS$_{2}$ (space group 187) for example, we calculate its energy
bands by \textsf{VASP} using the unit cell 1 shown in Fig. \ref{fig:MoS2}(a),
get \textsf{trace.txt }file by \textsf{vasp2trace}, and then determine
the LG IRs for all Bloch states in the bands by 
\begin{lstlisting}[backgroundcolor={\color{yellow!5!white}},mathescape=true]
|(* put trace.txt to the current working directory *)|
tr1=readVasp2trace["trace.txt"]; 
rep1=getBandRep[187, "", tr1];
\end{lstlisting}
in which \lstinline!rep1! is an association having keys \lstinline!"kpath"!,
\lstinline!"rep"!, and \lstinline!"kinfo"!, and the determined LG
IRs are contained in \lstinline!rep1["rep"]!. In this example, the
first k-point is $K$ $(\frac{2}{3},-\frac{1}{3},0)$ and the 10th
k-point is $K'$ $(-\frac{2}{3},\frac{1}{3},0)$. Then the LG IRs
of $K$ for seven valence bands 3\textendash 9 and four conduction
bands 10\textendash 13 can be extracted by \lstinline!rep1["rep"][[1, 3;;13]]!
whose result is shown in Fig. \ref{fig:MoS2}(b). The $\Gamma$ labels
$K_{1}\sim K_{6}$ of LG IRs in Fig. \ref{fig:MoS2}(b) should refer
to the character table in Fig. \ref{fig:MoS2}(c), from which it can
be seen that the characters of $C_{3}^{+}$ and $\sigma_{h}$ are
consistent with the data listed in the Tab. 2 of ref. \citep{Liu_Yao_2015_44_2643__Electronic}.
It should be pointed out that the LG IRs of $K'$, i.e. the results
of \lstinline!rep1["rep"][[10, 3;;13]]!, are \textit{seemingly} the
same with $K$ as shown by the ``True'' in Fig. \ref{fig:MoS2}(b).
However, the meanings of $K_{1}\sim K_{6}$ are different because
the LG IRs for $K'$ should refer to the character table in Fig. \ref{fig:MoS2}(d).
Taking the topmost valence band (the 9th band) as example, the character
of $\{C_{3}^{+}|000\}$ for the $K_{3}$ of $K$ is $e^{i\frac{2\pi}{3}}$
{[}c.f. Fig. \ref{fig:MoS2}(c){]}, but the character of $\{C_{3}^{+}|000\}$
for the $K_{3}$ of $K'$ is $e^{-i\frac{2\pi}{3}}$ {[}c.f. Fig.
\ref{fig:MoS2}(d){]}. The two characters are complex conjugates of
each other, which is consistent with the time reversal symmetry between
the states of $K$ and $K'$. Because $K'$ is not a BC standard k-point,
we should remember that it borrows the name $K$ from its related
BC standard k-point. In fact, if a non-BC k-point has its own name,
we can directly use this name in its labels of LG IRs, e.g. the $K_{1}\sim K_{6}$
in Fig. \ref{fig:MoS2}(d) are in fact $K_{1}'\sim K_{6}'$. 

It is noteworthy that the BC cell of a crystal is probably not unique.
Also taking monolayer MoS$_{2}$ for example, the three cells in Fig.
\ref{fig:MoS2}(a) are all BC cells, but they have different origins.
Consequently, the LG IR of a certain state is different for the three
different cells, because the rotation center is different, e.g. the
LG IRs of the topmost valence band and the lowest conduction band
(the 10th band) at $K$ for the three BC cells can be obtained as
follow
\begin{lstlisting}[backgroundcolor={\color{yellow!5!white}},mathescape=true]
|In[1]:=| rep1["rep"][[1, 9;;10]]//TableForm[#,TableDepth->2]&
        rep2["rep"][[1, 9;;10]]//TableForm[#,TableDepth->2]&
        rep3["rep"][[1, 9;;10]]//TableForm[#,TableDepth->2]&
|Out[1]=| {9,9}    -2.56193     1    {$^2$E$'$, K$_3$(1)}
        {10,10}  -0.886607    1    {A$'$, K$_1$(1)}
|Out[2]=| {9,9}    -2.56193     1    {A$'$, K$_1$(1)}
        {10,10}  -0.886609    1    {$^1$E$'$, K$_2$(1)}
|Out[3]=| {9,9}    -2.56188     1    {$^1$E$'$, K$_2$(1)}
        {10,10}  -0.886553    1    {$^2$E$'$, K$_3$(1)}
\end{lstlisting}
in which \lstinline!rep1!, \lstinline!rep2!, and \lstinline!rep3!
are the returned values of \lstinline!getBandRep! for BC cells 1,
2, and 3 in Fig. \ref{fig:MoS2}(a) respectively. The LG IRs for the
three BC cells are consistent with the eigenvalues of $C_{3}^{+}$
in the Tab. 2 of ref. \citep{Liu_Yao_2015_44_2643__Electronic} for
three different rotation centers. Therefore, if the primitive cell
is not given explicitly, we cannot say what is the LG IR of the topmost
valence band at $K$ (or any other state) for monolayer MoS$_{2}$.

\subsection{For non-BC cells}

For a non-BC primitive cell, it has to be converted to a BC cell and
its trace data also has to be converted accordingly before determining
LG IRs. To convert cells, we adopt the conventions used in the  package
\textsf{spglib}. In \textsf{spglib}, converting one cell to another
needs three ingredients, i.e. transformation matrix, rotation matrix,
and origin shift\citep{spglibdoc}. Concretely speaking, \textsf{spglib}
can convert any input cell with basic vectors $\bm{a}_{0}$, $\bm{b}_{0}$,
and $\bm{c}_{0}$ to an idealized standard cell with basic vectors
$\bm{a}_{s}'$, $\bm{b}_{s}'$, and $\bm{c}_{s}'$ through the relation
\begin{equation}
(\bm{a}_{s}',\bm{b}_{s}',\text{\ensuremath{\bm{c}_{s}')=R_{{\rm std}}(\bm{a}_{0},s\bm{b}_{0},\bm{c}_{0})}}P^{-1},\label{eq:Rabc0iP}
\end{equation}
in which $P$ and $R_{{\rm std}}$ are respectively the transformation
matrix and rotation matrix determined by \textsf{spglib}. The above
basic vectors are all column matrices, so both $(\bm{a}_{0},\bm{b}_{0},\bm{c}_{0})$
and $(\bm{a}_{s}',\bm{b}_{s}',\bm{c}_{s}')$ are $3\times3$ square
matrices and called ``basic-vector matrix''. In the conventions
of \textsf{spglib}, transformation matrix makes linear combination
of basic vectors to form new basic vectors but does not rotate the
crystal, while rotation matrix rotates the crystal and hence rotates
all basic vectors. Therefore, transformation matrix is always multiplied
on the right side of a basic-vector matrix, while rotation matrix
is always multiplied on the left side of a basic-vector matrix. Further
using the origin shift $\bm{p}$ determined by \textsf{spglib}, the
atomic coordinates and SG elements can be converted as follow
\begin{equation}
\bm{x}_{0}\ \ \ \ \rightarrow\ \ \ \ \bm{x}_{s}=P\bm{x}_{0}+\bm{p},
\end{equation}
\begin{equation}
\{R_{0}|\bm{v}_{0}\}\ \ \ \ \rightarrow\ \ \ \ \{R_{s}|\bm{v}_{s}\}=\{PR_{0}P^{-1}|P\bm{v}_{0}-R_{s}\bm{p}+\bm{p}\},
\end{equation}
in which $\bm{x}_{0}$ and $\{R_{0}|\bm{v}_{0}\}$ are respectively
the atomic coordinates and SG elements of the input cell, and $\bm{x}_{s}$
and $\{R_{s}|\bm{v}_{s}\}$ are respectively the atomic coordinates
and SG elements of the idealized standard cell of \textsf{spglib}.
In fact, the idealized standard cell of \textsf{spglib} is the conventional
cell consistent with the first setting of ITA\citep{ITA} and hence
can also be called ``ITA cell''. 

\begin{table}
\caption{Changes to BC-Tab. 3.7. All space groups that are not listed in this
table have the same data for the original BC-Tab. 3.7 and the adapted
BC-Tab. 3.7.\label{tab:BC3.7change}}

{\renewcommand{\arraystretch}{1.2}
\begin{centering}
\begin{tabular}{llll}
\hline 
No.$\ \ \ \ $ & Symbol$\ \ \ \ \ $ & Generators & $\bm{t}_{0}$\tabularnewline
\hline 
68 & $Ccca$ & \{$C_{2x}|000\},\ \{C_{2y}|000\},\ \{I|\frac{1}{2}\frac{1}{2}\frac{1}{2}\}$ & 0\tabularnewline
125 & $P4/nbm$ & \{$C_{4z}^{+}|\frac{1}{2}\frac{1}{2}0\},\ \{C_{2x}|000\},\ \{I|\frac{1}{2}\frac{1}{2}0\}$ & $\frac{1}{2}\bm{t}_{1}$\tabularnewline
 &  & \{$C_{4z}^{+}|000\},\ \{C_{2x}|000\},\ \{I|\frac{1}{2}\frac{1}{2}0\}$ & \tabularnewline
141 & $I4_{1}/amd$ & \{$C_{4z}^{+}|0\frac{1}{2}0\},\ \{C_{2x}|\frac{1}{2}\frac{1}{2}0\},\ \{I|\frac{1}{2}\frac{1}{2}0\}$ & $\frac{3}{8}\bm{t}_{1}+\frac{1}{8}\bm{t}_{2}+\frac{1}{4}\bm{t}_{3}$\tabularnewline
 &  & \{$C_{4z}^{+}|\frac{3}{4}\frac{1}{4}\frac{1}{2}\},\ \{C_{2x}|\frac{3}{4}\frac{1}{4}\frac{1}{2}\},\ \{I|\frac{3}{4}\frac{1}{4}\frac{1}{2}\}$ & \tabularnewline
142 & $I4_{1}/acd$ & \{$C_{4z}^{+}|\frac{1}{2}00\},\ \{C_{2x}|\frac{1}{2}\frac{1}{2}0\},\ \{I|000\}$ & $\frac{1}{8}\bm{t}_{1}+\frac{3}{8}\bm{t}_{2}-\frac{1}{4}\bm{t}_{3}$\tabularnewline
 &  & \{$C_{4z}^{+}|\frac{3}{4}\frac{1}{4}\frac{1}{2}\},\ \{C_{2x}|\frac{1}{4}\frac{3}{4}\frac{1}{2}\},\ \{I|\frac{3}{4}\frac{1}{4}\frac{1}{2}\}$ & \tabularnewline
155 & $R32$ & \{$C_{3}^{+}|000\},\ \{C_{21}'|000\}$ & 0\tabularnewline
160 & $R3m$ & \{$C_{3}^{+}|000\},\ \{\sigma_{d1}|000\}$ & 0\tabularnewline
161 & $R3c$ & \{$C_{3}^{+}|000\},\ \{\sigma_{d1}|\frac{1}{2}\frac{1}{2}\frac{1}{2}\}$ & 0\tabularnewline
166 & $R\bar{3}m$ & \{$S_{6}^{+}|000\},\ \{\sigma_{d1}|000\}$ & 0\tabularnewline
167 & $R\bar{3}c$ & \{$S_{6}^{+}|000\},\ \{\sigma_{d1}|\frac{1}{2}\frac{1}{2}\frac{1}{2}\}$ & 0\tabularnewline
178 & $P6_{1}22$ & \{$C_{6}^{+}|00\frac{1}{6}\},\ \{C_{21}'|000\}$ & $\frac{1}{4}\bm{t}_{3}$\tabularnewline
 &  & \{$C_{6}^{+}|00\frac{1}{6}\},\ \{C_{21}''|000\}$ & \tabularnewline
179 & $P6_{5}22$ & \{$C_{6}^{+}|00\frac{5}{6}\},\ \{C_{21}'|000\}$ & $\frac{1}{4}\bm{t}_{3}$\tabularnewline
 &  & \{$C_{6}^{+}|00\frac{5}{6}\},\ \{C_{21}''|000\}$ & \tabularnewline
182 & $P6_{3}22$ & \{$C_{6}^{+}|00\frac{1}{2}\},\ \{C_{21}'|000\}$ & $\frac{1}{4}\bm{t}_{3}$\tabularnewline
 &  & \{$C_{6}^{+}|00\frac{1}{2}\},\ \{C_{21}''|000\}$ & \tabularnewline
\hline 
\end{tabular}
\par\end{centering}
}

\end{table}

Next, we should convert the ITA cell to a BC cell with the aid of
BC-Tab. 3.7. BC-Tab. 3.7 gives the SG generators of each space group
and for some space groups there are two rows of data of which the
first-row generators are the ones used in the BC book and the second-row
generators are those used in the book \citep{IT1965} (referred to
as IT1965 hereafter) but based on the BC basic vectors. Then we call
the cell having the second-row SG generators ``second-row cell'',
and in this sense a BC cell is just a ``first-row cell''. If a space
group has only one row of generators in BC-Tab. 3.7, the term second-row
cell can also be used but has the same meaning as the first-row cell.
In fact, IT1965 can be considered as an earlier edition of ITA but
their SG settings have some differences. In order to convert the ITA
cell to a BC cell, BC-Tab. 3.7 has to be adapted and the changes are
listed in Tab. \ref{tab:BC3.7change}. So the BC-Tab. 3.7 with the
changes in Tab. \ref{tab:BC3.7change} is called ``the adapted BC-Tab.
3.7''. 

\begin{figure}[t]
\begin{centering}
\includegraphics[width=14cm]{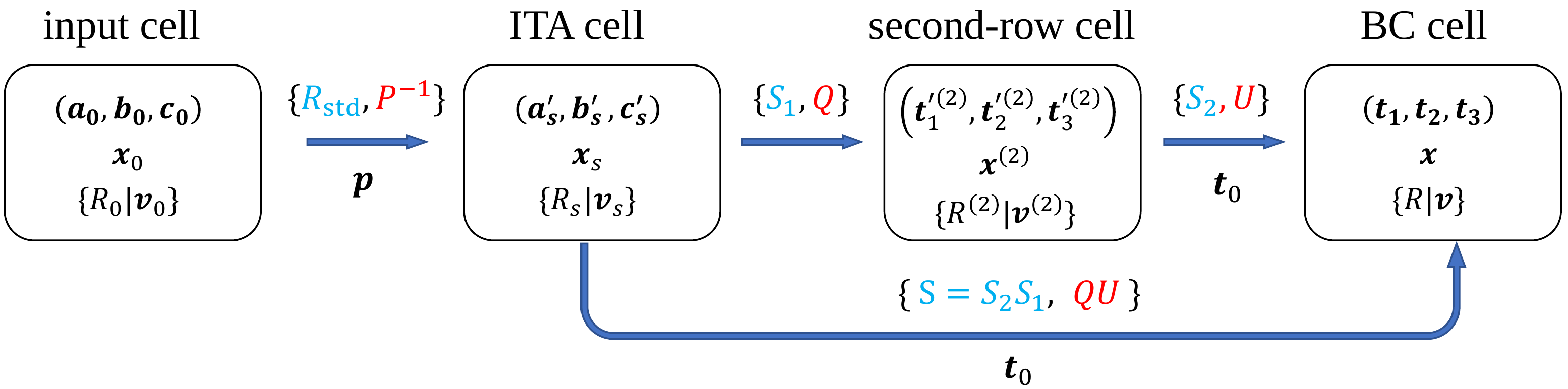}
\par\end{centering}
\caption{The procedure of converting the input cell to the BC cell. The transformation
matrices (red color) and rotation matrices (blue color) are shown
above the arrows, and the origin shifts (if exist) are shown below
the arrows.\label{fig:cell-convert}}

\end{figure}

According to the adapted BC-Tab. 3.7, we first convert the ITA cell
to a second-row cell, and then convert the second-row cell to a BC
cell. The first conversion has the transformation matrix $Q$, rotation
matrix $S_{1}$ and no origin shift, and the second conversion has
the transformation matrix $U$, rotation matrix $S_{2}$, and origin
shift $\bm{t}_{0}$. Suppose that the basic-vector matrices of the
second-row cell and the BC cell are $(\bm{t}_{1}'^{(2)},\bm{t}_{2}'^{(2)},\bm{t}_{3}'^{(2)})$
and $(\bm{t}_{1},\bm{t}_{2},\bm{t}_{3})$ respectively, that the atomic
coordinates of the two cells are $\bm{x}^{(2)}$ and $\bm{x}$ respectively,
and that the SG elements of the two cells are $\{R^{(2)}|\bm{v}^{(2)}\}$
and $\{R|\bm{v}\}$ respectively. Then the conversion from the ITA
cell to the second-row cell consists of the following relations
\begin{equation}
(\bm{t}_{1}'^{(2)},\bm{t}_{2}'^{(2)},\bm{t}_{3}'^{(2)})=S_{1}(\bm{a}_{s}',\bm{b}_{s}',\bm{c}_{s}')Q,\label{eq:S1abcspQ}
\end{equation}
\begin{equation}
\bm{x}_{s}\ \ \ \ \rightarrow\ \ \ \ \bm{x}^{(2)}=Q^{-1}\bm{x}_{s},
\end{equation}
\begin{equation}
\{R_{s}|\bm{v}_{s}\}\ \ \ \ \rightarrow\ \ \ \ \{R^{(2)}|\bm{v}^{(2)}\}=\{Q^{-1}R_{s}Q\,|\,Q^{-1}\bm{v}_{s}\},
\end{equation}
and the conversion from the second-row cell to the BC cell consists
of the following relations
\begin{equation}
(\bm{t}_{1},\bm{t}_{2},\bm{t}_{3})=S_{2}(\bm{t}_{1}'^{(2)},\bm{t}_{2}'^{(2)},\bm{t}_{3}'^{(2)})U,\label{eq:S2t123p2U}
\end{equation}
\begin{equation}
\bm{x}^{(2)}\ \ \ \ \rightarrow\ \ \ \ \bm{x}=U^{-1}\bm{x}^{(2)}+\bm{t}_{0},
\end{equation}
\begin{equation}
\{R^{(2)}|\bm{v}^{(2)}\}\ \ \ \ \rightarrow\ \ \ \ \{R|\bm{v}\}=\{U^{-1}R^{(2)}U\,|\,U^{-1}\bm{v}^{(2)}-R\bm{t}_{0}+\bm{t}_{0}\},\label{eq:Rv2toRv}
\end{equation}
where $\bm{t}_{0}$ is given in the adapted BC-Tab. 3.7.

\begin{table}[t]
\caption{The transformation matrix $Q$ and rotation matrix $S_{1}$ in Eq.
(\ref{eq:S1abcspQ}) for each Bravais lattice. ${\rm c}\gamma$ (${\rm s}\gamma$)
means $\cos\gamma$ ($\sin\gamma$) and $\gamma$ is the non-right
angle between basic vectors of monoclinic lattices given in BC-Tab.
3.1.\label{tab:QS1}}

{\renewcommand{\arraystretch}{1.2}
\begin{centering}
\begin{tabular}{llllll}
\hline 
\Gape{%
\parbox[c]{4em}{%
Bravais lattice%
}} & $Q$ & $S_{1}$ & %
\parbox[c]{4em}{%
Bravais lattice%
} & $Q$ & $S_{1}$\tabularnewline
\hline 
\begin{lstlisting}
TricPrim
\end{lstlisting}
 & \Gape{$\begin{bmatrix}1 & 0 & 0\\
0 & 1 & 0\\
0 & 0 & 1
\end{bmatrix}$} & $\begin{bmatrix}1 & 0 & 0\\
0 & 1 & 0\\
0 & 0 & 1
\end{bmatrix}$ & 
\begin{lstlisting}
TetrPrim
\end{lstlisting}
 & $\begin{bmatrix}1 & 0 & 0\\
0 & 1 & 0\\
0 & 0 & 1
\end{bmatrix}$ & $\begin{bmatrix}1 & 0 & 0\\
0 & 1 & 0\\
0 & 0 & 1
\end{bmatrix}$\tabularnewline
\begin{lstlisting}
MonoPrim
\end{lstlisting}
 & \Gape{$\begin{bmatrix}0 & 1 & 0\\
0 & 0 & 1\\
1 & 0 & 0
\end{bmatrix}$} & $\begin{bmatrix}{\rm s}\gamma & 0 & -{\rm c}\gamma\\
-{\rm c}\gamma & 0 & -{\rm s}\gamma\\
0 & 1 & 0
\end{bmatrix}$ & 
\begin{lstlisting}
TetrBody
\end{lstlisting}
 & $\begin{bmatrix}-\frac{1}{2} & \frac{1}{2} & \frac{1}{2}\\
\frac{1}{2} & -\frac{1}{2} & \frac{1}{2}\\
\frac{1}{2} & \frac{1}{2} & -\frac{1}{2}
\end{bmatrix}$ & $\begin{bmatrix}1 & 0 & 0\\
0 & 1 & 0\\
0 & 0 & 1
\end{bmatrix}$\tabularnewline
\begin{lstlisting}
MonoBase
\end{lstlisting}
 & \Gape{$\begin{bmatrix}0 & \frac{1}{2} & \frac{1}{2}\\
0 & -\frac{1}{2} & \frac{1}{2}\\
1 & 0 & 0
\end{bmatrix}$} & $\begin{bmatrix}{\rm s}\gamma & 0 & -{\rm c}\gamma\\
-{\rm c}\gamma & 0 & -{\rm s}\gamma\\
0 & 1 & 0
\end{bmatrix}$ & 
\begin{lstlisting}
TrigPrim
\end{lstlisting}
 & $\begin{bmatrix}\frac{2}{3} & -\frac{1}{3} & -\frac{1}{3}\\
\frac{1}{3} & \frac{1}{3} & -\frac{2}{3}\\
\frac{1}{3} & \frac{1}{3} & \frac{1}{3}
\end{bmatrix}$ & $\begin{bmatrix}-\frac{1}{2} & \frac{\sqrt{3}}{2} & 0\\
-\frac{\sqrt{3}}{2} & -\frac{1}{2} & 0\\
0 & 0 & 1
\end{bmatrix}$\tabularnewline
\begin{lstlisting}
OrthPrim
\end{lstlisting}
 & \Gape{$\begin{bmatrix}0 & 1 & 0\\
-1 & 0 & 0\\
0 & 0 & 1
\end{bmatrix}$} & $\begin{bmatrix}1 & 0 & 0\\
0 & 1 & 0\\
0 & 0 & 1
\end{bmatrix}$ & 
\begin{lstlisting}
HexaPrim
\end{lstlisting}
 & $\begin{bmatrix}1 & 0 & 0\\
0 & 1 & 0\\
0 & 0 & 1
\end{bmatrix}$ & $\begin{bmatrix}0 & 1 & 0\\
-1 & 0 & 0\\
0 & 0 & 1
\end{bmatrix}$\tabularnewline
\begin{lstlisting}
OrthBase
\end{lstlisting}
 & \Gape{$\begin{bmatrix}0 & 1 & 0\\
-\frac{1}{2} & 0 & \frac{1}{2}\\
\frac{1}{2} & 0 & \frac{1}{2}
\end{bmatrix}$} & $\begin{bmatrix}1 & 0 & 0\\
0 & 1 & 0\\
0 & 0 & 1
\end{bmatrix}$ & 
\begin{lstlisting}
CubiPrim
\end{lstlisting}
 & $\begin{bmatrix}1 & 0 & 0\\
0 & 1 & 0\\
0 & 0 & 1
\end{bmatrix}$ & $\begin{bmatrix}1 & 0 & 0\\
0 & 1 & 0\\
0 & 0 & 1
\end{bmatrix}$\tabularnewline
\begin{lstlisting}
OrthBody
\end{lstlisting}
 & \Gape{$\begin{bmatrix}\frac{1}{2} & -\frac{1}{2} & \frac{1}{2}\\
\frac{1}{2} & -\frac{1}{2} & -\frac{1}{2}\\
\frac{1}{2} & \frac{1}{2} & -\frac{1}{2}
\end{bmatrix}$} & $\begin{bmatrix}1 & 0 & 0\\
0 & 1 & 0\\
0 & 0 & 1
\end{bmatrix}$ & 
\begin{lstlisting}
CubiFace
\end{lstlisting}
 & $\begin{bmatrix}0 & \frac{1}{2} & \frac{1}{2}\\
\frac{1}{2} & 0 & \frac{1}{2}\\
\frac{1}{2} & \frac{1}{2} & 0
\end{bmatrix}$ & $\begin{bmatrix}1 & 0 & 0\\
0 & 1 & 0\\
0 & 0 & 1
\end{bmatrix}$\tabularnewline
\begin{lstlisting}
OrthFace
\end{lstlisting}
 & \Gape{$\begin{bmatrix}\frac{1}{2} & 0 & \frac{1}{2}\\
0 & -\frac{1}{2} & -\frac{1}{2}\\
\frac{1}{2} & \frac{1}{2} & 0
\end{bmatrix}$} & $\begin{bmatrix}1 & 0 & 0\\
0 & 1 & 0\\
0 & 0 & 1
\end{bmatrix}$ & 
\begin{lstlisting}
CubiFace
\end{lstlisting}
 & $\begin{bmatrix}-\frac{1}{2} & \frac{1}{2} & \frac{1}{2}\\
\frac{1}{2} & -\frac{1}{2} & \frac{1}{2}\\
\frac{1}{2} & \frac{1}{2} & -\frac{1}{2}
\end{bmatrix}$ & $\begin{bmatrix}1 & 0 & 0\\
0 & 1 & 0\\
0 & 0 & 1
\end{bmatrix}$\tabularnewline
\hline 
\end{tabular}
\par\end{centering}
}
\end{table}

To sum up briefly, any input cell can be converted to BC cell via
two intermediate cells and the procedure includes three steps: firstly
input cell to ITA cell, then ITA cell to second-row cell, and lastly
second-row cell to BC cell, as shown in Fig. \ref{fig:cell-convert}.
In the first step, $R_{{\rm std}}$, $P$, and $\bm{p}$ are all determined
by \textsf{spglib}; in the second step, $Q$ and $S_{1}$ are listed
in Tab. \ref{tab:QS1} according to the Bravais lattice of space group;
and in the last step, $U$ and $S_{2}$ are listed in Tab. \ref{tab:US2}
and $\bm{t}_{0}$ is given in the last column of the adapted BC-Tab.
3.7. Integrating the three steps from Eq. (\ref{eq:Rabc0iP}) to Eq.
(\ref{eq:Rv2toRv}), the integrated conversion from the input cell
to the BC cell consists of the following relations
\begin{equation}
(\bm{t}_{1},\bm{t}_{2},\bm{t}_{3})=S_{2}S_{1}R_{{\rm std}}(\bm{a}_{0},\bm{b}_{0},\bm{c}_{0})P^{-1}QU,
\end{equation}
\begin{equation}
\bm{x}_{0}\ \ \ \ \rightarrow\ \ \ \ \bm{x}=U^{-1}Q^{-1}(P\bm{x}_{0}+\bm{p})+\bm{t}_{0},\label{eq:x0tox}
\end{equation}
\begin{equation}
\{R_{0}|\bm{v}_{0}\}\ \ \ \ \rightarrow\ \ \ \ \{R|\bm{v}\}\begin{cases}
R=U^{-1}Q^{-1}PR_{0}P^{-1}QU\\
\bm{v}=U^{-1}Q^{-1}(P\bm{v}_{0}-PR_{0}P^{-1}\bm{p}+\bm{p})-R\bm{t}_{0}+\bm{t}_{0}
\end{cases}.\label{eq:R0v0toRv}
\end{equation}

\begin{table}[ph]
\caption{The transformation matrix $U$ and rotation matrix $S_{2}$ in Eq.
(\ref{eq:S2t123p2U}) for the space groups with an $*$ in the last
column of the adapted BC-Tab. 3.7. The BC book uses orientations different
from the default ones in ITA for these space groups, and the orientations
used are given here in the second column (refer to ITA for the meanings
of the orientations). For the space groups not listed in this table,
both $U$ and $S_{2}$ are identity matrix.\label{tab:US2}}

{\renewcommand{\arraystretch}{1.2}
\begin{centering}
\begin{tabular}{llccc}
\hline 
\multicolumn{2}{c}{Space Group} & orientation & $U$ & $S_{2}$\tabularnewline
\hline 
$17\ (P222_{1})$ & $19\ (P2_{1}2_{1}2_{1})$ & \multirow{8}{*}{\Gape[12.5em]{$\bm{b\bar{a}c}$}} & \multirow{4}{*}{\Gape{$\begin{bmatrix}\begin{array}{ccc}
0 & -1 & 0\\
1 & 0 & 0\\
0 & 0 & 1
\end{array}\end{bmatrix}$}} & \multirow{8}{*}{\Gape[11em]{$\begin{bmatrix}0 & 1 & 0\\
-1 & 0 & 0\\
0 & 0 & 1
\end{bmatrix}$}}\tabularnewline
$28\ (Pma2)$ & $29\ (Pca2_{1})$ &  &  & \tabularnewline
$31\ (Pmn2_{1})$ & $33\ (Pna2_{1})$ &  &  & \tabularnewline
$53\ (Pmna)$ & $61\ (Pbca)$ &  &  & \tabularnewline
$36\ (Cmc2_{1})$ &  &  & \Gape{$\begin{bmatrix}\begin{array}{ccc}
0 & -1 & 0\\
1 & 0 & 0\\
0 & 0 & 1
\end{array}\end{bmatrix}$} & \tabularnewline
$46\ (Ima2)$ &  &  & \Gape{$\begin{bmatrix}\begin{array}{ccc}
0 & 1 & 0\\
0 & 1 & -1\\
-1 & 1 & 0
\end{array}\end{bmatrix}$} & \tabularnewline
$70\ (Fddd)$ &  &  & \Gape{$\begin{bmatrix}\begin{array}{ccc}
1 & 1 & 1\\
0 & 0 & -1\\
-1 & 0 & 0
\end{array}\end{bmatrix}$} & \tabularnewline
$122\ (I\bar{4}2d$) &  &  & \Gape{$\begin{bmatrix}\begin{array}{ccc}
0 & 1 & 0\\
0 & 1 & -1\\
-1 & 1 & 0
\end{array}\end{bmatrix}$} & \tabularnewline
\hline 
\begin{minipage}[c]{2cm}%
\mbox{}\\

38 $(Amm2)$\\[-8pt]

40 $(Ama2)$

\mbox{}\\[-1em]%
\end{minipage} & %
\begin{minipage}[c]{2cm}%
\mbox{}\\

39 $(Abm2)$\\[-8pt]

41 $(Aba2)$

\mbox{}\\[-1em]%
\end{minipage} & \multirow{2}{*}{\Gape[2.5em]{$\bm{bca}$}} & \multirow{1}{*}{\Gape[-2em]{$\begin{bmatrix}\begin{array}{ccc}
-1 & 0 & 0\\
0 & 0 & 1\\
0 & 1 & 0
\end{array}\end{bmatrix}$}} & \multirow{2}{*}{\Gape[1em]{$\begin{bmatrix}0 & 1 & 0\\
0 & 0 & 1\\
1 & 0 & 0
\end{bmatrix}$}}\tabularnewline
$57\ (Pbcm)$ &  &  & \Gape{$\begin{bmatrix}\begin{array}{ccc}
0 & -1 & 0\\
0 & 0 & 1\\
-1 & 0 & 0
\end{array}\end{bmatrix}$} & \tabularnewline
\hline 
$51\ (Pmma)$ & $54\ (Pcca)$ & $\bm{\bar{c}ba}$ & $\begin{bmatrix}\begin{array}{ccc}
1 & 0 & 0\\
0 & 0 & 1\\
0 & -1 & 0
\end{array}\end{bmatrix}$ & \Gape{$\begin{bmatrix}0 & 0 & -1\\
0 & 1 & 0\\
1 & 0 & 0
\end{bmatrix}$}\tabularnewline
\hline 
$52\ (Pnna)$ & $60\ (Pbcn)$ & $\bm{a}\bar{\bm{c}}\bm{b}$ & $\begin{bmatrix}\begin{array}{ccc}
0 & 0 & -1\\
0 & 1 & 0\\
1 & 0 & 0
\end{array}\end{bmatrix}$ & \Gape{$\begin{bmatrix}1 & 0 & 0\\
0 & 0 & -1\\
0 & 1 & 0
\end{bmatrix}$}\tabularnewline
\hline 
\end{tabular}
\par\end{centering}
}
\end{table}

From Eqs. (\ref{eq:x0tox}) and (\ref{eq:R0v0toRv}) we can see that
in the conversion of atomic coordinates and SG elements no rotation
matrices ($R_{{\rm std}}$, $S_{1}$, or $S_{2})$ are used. This
is because the rotation of crystal rotates the basic vectors and atomic
positions simultaneously but does not change the fractional coordinates
of atoms. On the contrary, the conversion of the spin rotation matrices
for double space groups uses only the rotation matrices, i.e.
\begin{equation}
\tilde{R}=\tilde{S}\tilde{R}_{{\rm std}}\tilde{R}_{0}\tilde{R}_{{\rm std}}^{-1}\tilde{S}^{-1},
\end{equation}
in which $\tilde{R}_{0}$ and $\tilde{R}$ are the spin rotation matrices
of the input cell and the BC cell respectively, and $\tilde{R}_{{\rm std}}$
and $\tilde{S}$ are the SU(2) spin rotation matrices of the corresponding
O(3) rotation matrices $R_{{\rm std}}$ and $S$ $(=S_{2}S_{1})$
respectively. $\tilde{S}$ is determined by $S$ through $\tilde{S}=\exp(-i\omega\bm{n}\cdot\bm{\sigma}/2)$
in which $\omega$ and $\bm{n}$ are the rotation angle and the unit
direction of the rotation axis of $S$ respectively and $\bm{\sigma}$
is the vector of Pauli matrices, and so is $\tilde{R}_{{\rm std}}.$ 

Using the cell conversion method mentioned above, the data in a\textsf{
trace.txt }file from a non-BC cell can be converted to the data for
a BC cell by
\begin{lstlisting}[backgroundcolor={\color{yellow!5!white}}]
convTraceToBC[sgno, traceData, P, p, stdR]    
\end{lstlisting}
in which \lstinline!traceData! is the returned value of \lstinline!readVasp2trace!
for a non-BC cell, and \lstinline!P!, \lstinline!p!, and \lstinline!stdR!
are respectively the transformation matrix $P$, the origin shift
$\bm{p}$, and the rotation matrix $R_{{\rm std}}$ determined by
\textsf{spglib}. Note that \lstinline!stdR! is needed only when spin-orbit
coupling is considered for the trace data. This conversion can also
be done automatically by 
\begin{lstlisting}[backgroundcolor={\color{yellow!5!white}}]
autoConvTraceToBC[poscarFile, traceData]   
\end{lstlisting}
in which \lstinline!poscarFile! is the file name of the \textsf{POSCAR}
file for the non-BC cell. In fact, \lstinline!autoConvTraceToBC!
first calls the function \lstinline!readPOSCAR! to read the non-BC
\textsf{POSCAR} file and then calls the function \lstinline!spglibGetSym!
which calls the python interface of the package \textsf{spglib} externally
to determine $P$, $\bm{p}$, and $R_{{\rm std}}$. After the trace
data are converted they belong to a BC cell and can be directly used
by \lstinline!getBandRep! to determine LG IRs.

\section{Correspondence of LG IR labels between BCS and BC conventions}

\begin{figure}[th]
\begin{centering}
\includegraphics[width=14cm]{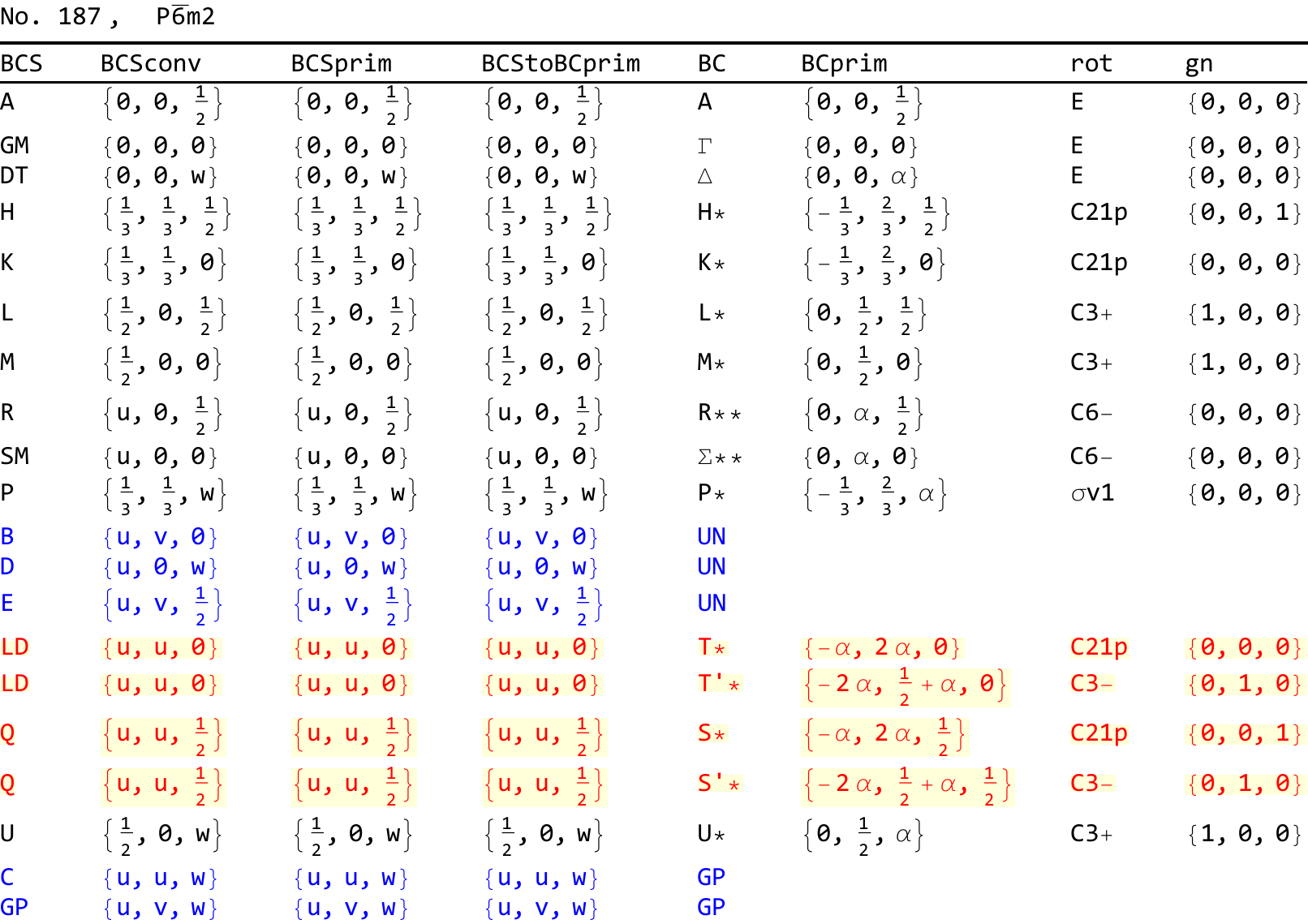}
\par\end{centering}
\caption{The output of \lstinline[backgroundcolor={\color{yellow!5!white}}]!showKptBCStoBC[187]!
which gives the correspondence of k-point coordinates between BCS
and BC conventions. The first column is the BCS k-points. The second
and third columns are the BCS k-point coordinates for conventional
cell and primitive cell respectively. The 4th column is the k-point
coordinates for a BC cell which are converted from the BCS k-point
coordinates, i.e. the $\protect\vk$ in Eq. (\ref{eq:k0tok}). The
5th column is the BC k-points whose BC standard coordinates, namely
$\protect\vk_{{\rm BC}},$ are given in the 6th column. The relation
between $\protect\vk$ and $\protect\vk_{{\rm BC}}$ is $\protect\vk=S\protect\vk_{{\rm BC}}+\bm{g}_{n}$
in which $S$ is the rotation given in the 7th column and $\bm{g}_{n}$
is the reciprocal lattice vector given in the 8th column. The k-points
with $*$ and $**$ in the 5th column are respectively of type II
and type III as defined in subsection \ref{subsec:LGIR-at-any-k}.
Blue color highlights the k-points of type IV (GP) and type V (UN).
Red color highlights the k-points which have different names in BCS
and BC conventions. Yellow background highlights the cases in which
one BCS k-point may be identified as two BC k-points. \label{fig:kptBCStoBC187}}

\end{figure}

\begin{figure}[th]
\begin{centering}
\includegraphics[width=12cm]{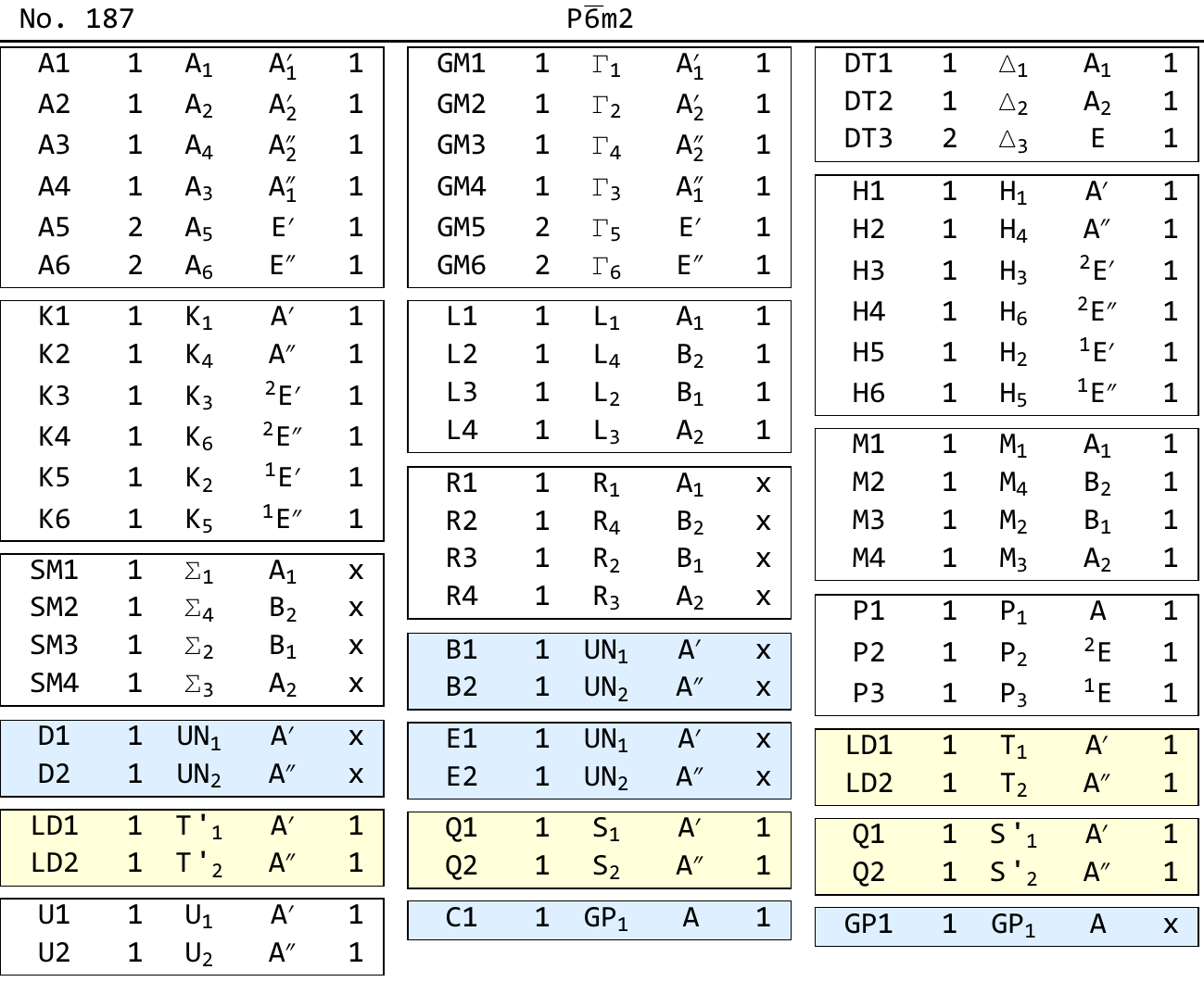}
\par\end{centering}
\caption{The output of \lstinline[backgroundcolor={\color{yellow!5!white}}]!showKrepBCStoBC[187]!
which gives the correspondence of LG IR labels between BCS and BC
conventions. The first column is the $\Gamma$ labels for BCS LG IRs.
The second column is the dimensions of the LG IRs. The third column
is the $\Gamma$ labels for BC LG IRs. The 4th column is the extended
Mulliken labels. The 5th column is the realities for corresponding
SG IRs. Blue background highlights the k-points of type IV (GP) and
type V (UN). Yellow background highlights the cases in which one BCS
k-point may be identified as two BC k-points.\label{fig:KrepBCStoBC187}}

\end{figure}

With the aid of \textsf{irvsp}, we have obtained all the data of LG
IRs in BCS convention. Based on these BCS data we can make correspondence
of LG IR labels between BCS and BC conventions. To achieve this, the
coordinates of all the k-points defined in BCS convention have to
be converted to the coordinates in BC convention. The conversion of
k-point coordinates can be done through the equation
\begin{equation}
\bm{k}=(P^{-1}QU)^{T}\vk_{0},\label{eq:k0tok}
\end{equation}
in which $\vk_{0}$ is the k-point defined by a BCS cell, namely the
input cell, $\vk$ is the k-point defined by the converted BC cell,
and $P$, $Q$, and $U$ are the transformation matrices of the aforementioned
cell conversion. Then $\vk$ is processed by \lstinline!identifyBCHSKptBySG!
to find its relation to the BC standard k-point $\vk_{{\rm BC}}$.
The correspondence of k-point coordinates between BCS and BC conventions
can be tabulated by the function 
\begin{lstlisting}[backgroundcolor={\color{yellow!5!white}}]
showKptBCStoBC[sgno, BZtype]
\end{lstlisting}
for the space group of number \lstinline!sgno!, in which \lstinline!BZtype!
is optional if it is \lstinline!""! or \lstinline!"a"!. An example
is shown in Fig. \ref{fig:kptBCStoBC187} for space group $P\bar{6}m2$
(No. 187) where the second or third column corresponds to $\vk_{0}$
and the 4th and 6th columns correspond to $\vk$ and $\vk_{{\rm BC}}$
respectively. After the correspondence of k-points is clear, the correspondence
of LG IR labels can be made by first building trace data containing
all the BCS LG IRs and then determining their BC LG IRs via \lstinline!getBandRep!.
In this procedure, it has to be pointed out that it is the complex
conjugates of the BCS characters that can correspond to the BC ones.
A typical example is that the character of a pure translation $\{E|t_{1}t_{2}t_{3}\}$
is $e^{i2\pi(ut_{1}+vt_{2}+wt_{3})}$ for the k-point $(u,v,w)$ in
BCS convention while it should be $e^{-i2\pi(ut_{1}+vt_{2}+wt_{3})}$
in BC convention. The final correspondence of LG IR labels can be
tabulated by the function \lstinline!showKrepBCStoBC! for either
single-valued or double-valued LG IRs, 
\begin{lstlisting}[backgroundcolor={\color{yellow!5!white}}]
showKrepBCStoBC[sgno, BZtype]               (* for sigle-valued LG IRs *)
showKrepBCStoBC[sgno, BZtype, "DSG"->True]  (* for double-valued LG IRs *)
\end{lstlisting}
in which \lstinline!BZtype! is also optional if it is \lstinline!""!
or \lstinline!"a"!. The example for single-valued LG IRs of the space
group $P\bar{6}m2$ is shown in Fig. \ref{fig:KrepBCStoBC187}. 

\section{Conclusions}

During the development of the package \textsf{SpaceGroupIrep}, we
found some typos in the BC book. The fixed typos are given in the
supplementary material. For quick reference, the elements of each
space group in BC convention are listed in the supplementary material.
In addition, the correspondences of k-points and LG IR labels between
BCS and BC conventions for all the 230 space groups and all possible
types of BZs are also given in the supplementary material. 

In conclusion, we have developed a program package called \textsf{SpaceGroupIrep}
in the Mathematica language for space groups and their IRs in BC convention.
This package digitizes many tables in the BC book, especially the
huge tables BC-Tabs. 5.1, 5.7, and 6.13, and it provides tens of functions
to manipulate these data. In this package, there are functions which
can get the elements of a space group, a little group, a Herring little
group, or a central extension of little co-group and functions which
can calculate the multiplication of the elements. There are functions
which can get and show the LG IRs (SG IRs) of any k-point (k-star)
for both single-valued and double-valued IRs. There are functions
which can calculate and show the decomposition of the direct product
of SG IRs for any two k-stars. There are functions which can determine
the LG IRs of Bloch band states in BC convention from the \textsf{trace.txt}
file produced by \textsf{vasp2trace} and they work for any primitive
cell because there are functions which can convert any input cells
to BC cells. And there are also functions which give the correspondence
of k-points and LG IR labels between BCS and BC conventions. In addition
to the main functions mentioned above, there are other useful functions
such as \lstinline!showBZDemo! (showing the rotatable BZ and HS k-points
and k-lines), \lstinline!rotAxisAngle! (finding the rotation axis
and rotation angle of an O(3) matrix), and \lstinline!generateGroup!
(obtaining all group elements according to its generators and multiplication).
Detailed information for each function can be obtained by the Mathematica
build-in function \lstinline!Information!, e.g. \lstinline!generateGroup//Information!
or just \lstinline!?generateGroup!. In a word, the Mathematica package
\textsf{SpaceGroupIrep} is a very useful database and tool set for
both studying the representation theory of space group and applying
them in research such as analyzing band topology or determining selection
rules.

\section*{Acknowledgments}

GBL acknowledges the support by the National Key R\&D Program of China
(Grant No. 2017YFB0701600). ZZ acknowledges the support by China Postdoctoral
Science Foundation (Grant No. 2020M670106). YY acknowledges the support
by the National Key R\&D Program of China (Grant Nos. 2020YFA0308800
and 2016YFA0300600), the NSF of China (Grants No. 11734003), and the
Strategic Priority Research Program of Chinese Academy of Sciences
(Grant No. XDB30000000).

\bibliographystyle{elsarticle-num-names}
\bibliography{SpaceGroupIrep}

\includepdf[pages=-]{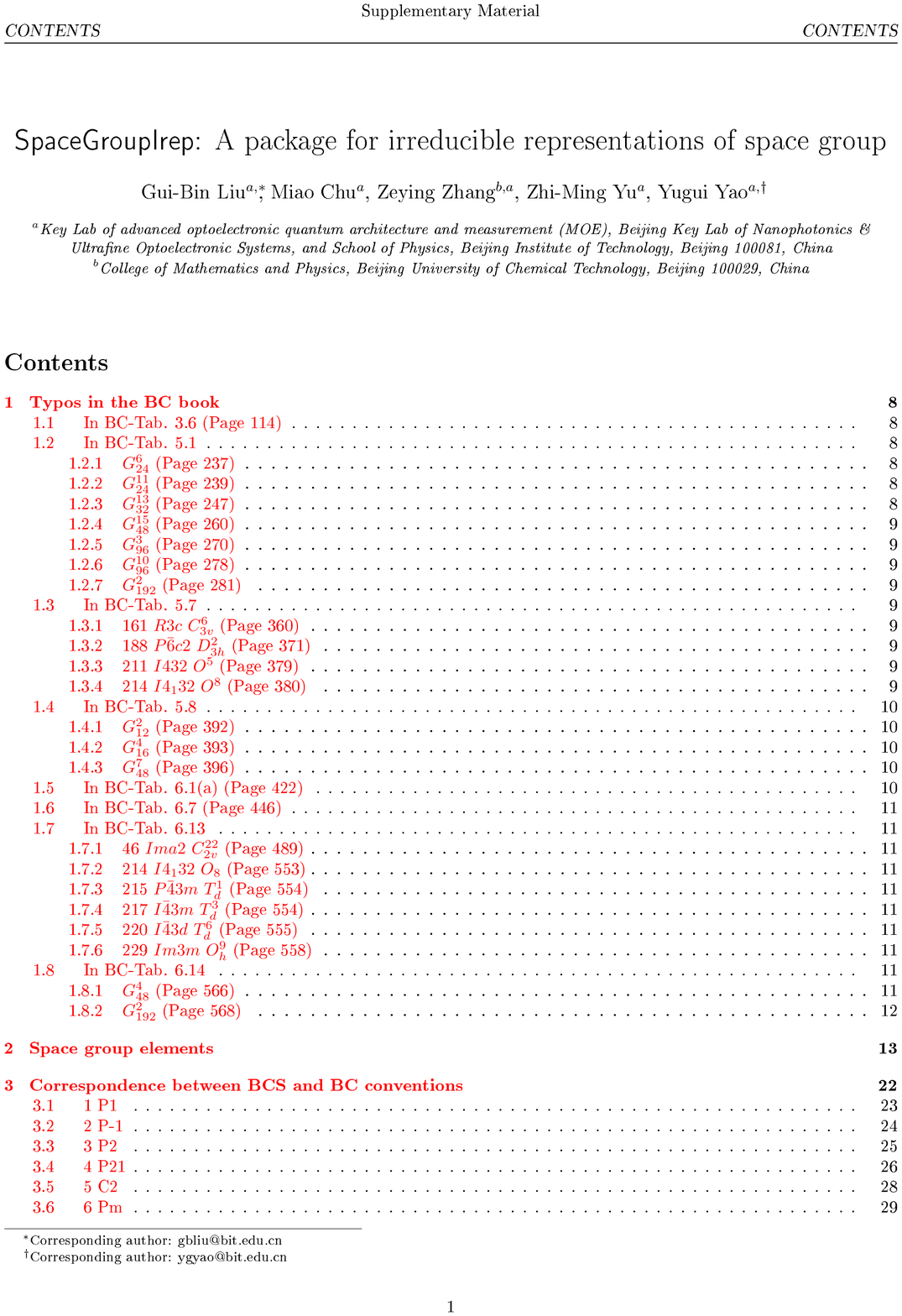}

\end{document}